\documentclass[superscriptaddress,twocolumn,secnumarabic,
nobibnotes,aps,prd,showpacs,nofootinbib]{revtex4}
\usepackage{graphicx}
\usepackage{epsf}
\usepackage{bm}
\usepackage{amsmath}
\usepackage{amsfonts}
\usepackage{amssymb}
\usepackage{epstopdf}
\usepackage{natbib}
\usepackage{color}
\usepackage{hyperref}
\usepackage{cleveref}

\providecommand{\U}[1]{\protect\rule{.1in}{.1in}}
\newcommand{\be}{\begin{equation}}
\newcommand{\ee}{\end{equation}}

\newcommand{\mincir}{\raise
-3.truept\hbox{\rlap{\hbox{$\sim$}}\raise4.truept\hbox{$<$}\ }}
\newcommand{\magcir}{\raise
-3.truept\hbox{\rlap{\hbox{$\sim$}}\raise4.truept\hbox{$>$}\ }}

\begin{document}

\title{Astronomical bounds on a cosmological model allowing a general interaction in the dark sector}

\author{Supriya Pan\footnote{\textit{Present Address: Department of Mathematics, Raiganj Surendranath Mahavidyalaya, Sudarshanpur, Raiganj, West Bengal 733134, India.}}}
\email{span@research.jdvu.ac.in}
\affiliation{Department of Physical Sciences, Indian Institute of Science Education and Research Kolkata, \\ Mohanpur, West Bengal 741246, India}

\author{Ankan Mukherjee}
\email{ankanju@iisermohali.ac.in}
\affiliation{Department of Physical Sciences, Indian Institute of Science Education and Research Kolkata, \\  Mohanpur, West Bengal 741246, India}
\affiliation{Department of Physical Sciences, Indian Institute of Science Education and Research Mohali, \\  Sector 81, Mohali, Punjab 140306, India}

\author{Narayan Banerjee}
\email{narayan@iiserkol.ac.in}
\affiliation{Department of Physical Sciences, Indian Institute of Science Education and Research Kolkata, \\ Mohanpur, West Bengal 741246, India}

\begin{abstract}
Non-gravitational interaction between two barotropic dark fluids, namely the pressureless dust and the dark energy in a spatially flat Friedmann-Lema\^{i}tre-Robertson-Walker model has been discussed. It is shown that for the interactions which are linear in terms the energy densities of the dark components and their first order derivatives, the net energy density is governed by a second order differential equation with constant coefficients. Taking a generalized interaction, which includes a number of already known interactions as special cases, the dynamics of the universe is described for three types of the dark energy equation of state, namely that of interacting quintessence, interacting vacuum energy density and  interacting phantom. The models have been constrained  using the standard cosmological probes, Supernovae type Ia data from joint light curve analysis and the observational Hubble parameter data. Two geometric tests, the cosmographic studies and the $Om$ diagnostic have been invoked so as to ascertain the behaviour of the present model vis-a-vis the $\Lambda$-cold dark matter model. We further discussed the interacting scenarios taking into account the thermodynamic considerations. 
\end{abstract}

\maketitle

\section{Introduction}
\label{Intro}

According to very recent observations \cite{Planck2015}, the dark matter and the dark energy are the two main sources of the total energy content of the universe. The dark matter is pressureless whereas dark energy is supposed to have a large negative pressure which fuels the expansion of the universe in an accelerated fashion. This picture of the universe can successfully be reproduced by the $\Lambda$CDM cosmology where $\Lambda$, the constant vacuum energy density, serves as the dark energy. Though $\Lambda$CDM cosmology is well consistent with most of the observational data, it has some serious issues, such as the {\it fine tuning problem} \cite{Weinberg1989, paddy}. The magnitude of $\Lambda$, according to observations, is very low, and that is sufficient for driving this current accelerating phase. Whereas, the theoretically predicted value of $\Lambda$, estimated from the quantum theory of fields, is enormously higher and consequently there is a huge discrepancy of order $10^{121}$ between the observed and theoretical estimations of $\Lambda$. So, the time dependent dark energy \cite{CDS1998} has been invoked as a plausible way out from the issues related to the cosmological constant (for a review of such time dependent dark energy, see \cite{varun, CST, AT, Bamba:2012cp}). However, the dynamical dark energy models have issues of their own, such as the {\it cosmic coincidence problem} \cite{Zlatev},  why the magnitudes of dark energy and dark matter are of same order today though they evolve independently with the evolution of the universe. The cosmological constant  $\Lambda$,  inherits the same problem although it does not evolve with the cosmic time. As a  result, several alternative cosmological models have been introduced in the last couple of decades with an aim to search for the dynamics of the universe \cite{CST, AT, Bamba:2012cp} where such issues can be resolved. The scenario of non-gravitational interaction between dark matter and dark energy is one of such alternative models, which is the main subject  interest of the present work. 

If the dark matter and the dark energy interact with each other, it is obvious that the total energy density should be conserved instead of their individual conservation. That means the balance equation for the total dark sector of the universe reads $\nabla_{\mu}(T_{m}^{\mu \nu}+ T_{d}^{\mu \nu})= 0$, where $T_{m}^{\mu \nu}$, $T_{d}^{\mu \nu}$, respectively denote the energy-momentum tensors of dark matter and dark energy, and consequently, this balance equation introduces a function $Q$ as, $\nabla_{\mu} T_{m}^{\mu \nu}= - \nabla_{\mu} T_{d}^{\mu \nu} = Q$. The main goal is to describe the dynamics of the universe in presence of such unknown function $Q$, known as the interaction function. If $Q=0$, the two matter components conserve themselves individually, and one gets back the non-interacting model. Thus, the interaction in the dark sector is indeed a more general scenario to unveil the dynamics of the universe. 

 The interaction in the dark sector was motivated so as to provide a resolution of the cosmic coincidence problem \cite{Wetterich-ide1, Amendola-ide1} (see also \cite{Amendola-ide2, Pavon:2005yx, delCampo:2008sr, delCampo:2008jx}) and there has been a lot of work done in this connection 
\cite{Billyard-ide1, Zimdahl-ide1, Amendola-ide2, Herrera-ide1, Chimento-ide1, Pavon:2005yx, He1, Quartin1, delCampo:2008sr, Maartens2008, Maartens2009, delCampo:2008jx, Valiviita:2009nu, Chimento2010, Clemson:2011an, Chen:2011cy, Pan2013, Faraoni:2014vra, Pan:2012ki, AP, Duniya:2015nva, Valiviita:2015dfa, delCampo:2015vha, Pan:2016ngu, Mukherjee:2016shl, Pan2017, Odintsov:2017icc, Yang:2017yme, DiValentino:2017iww, Yang:2017zjs, Yang:2017ccc, Santos:2017bqm, Kumar:2017bpv}. Recently, a series of investigations reported that the current observational data favour an interaction in the dark sector \cite{Salvatelli2014, Weiqiang1, Weiqiang2, yang:2014vza, Nunes:2016dlj, Kumar:2016zpg, Bruck2016, Weiqiang3, Yang:2017yme, DiValentino:2017iww, Yang:2017zjs, Yang:2017ccc, Santos:2017bqm, Kumar:2017bpv}. In fact, the interacting scenario can also take care of the issues connected to the local value of the Hubble parameter \cite{Kumar:2016zpg, PTram2016, DiValentino:2017iww}. However, it is also shown that a phantom universe may result through the dark matter dark energy interaction \cite{Sadjadi:2006qb, Pan:2014afa, Nunes:2016dlj, Yang:2017yme, Yang:2017zjs} without invoking any scalar field theory with negative kinetic term \cite{Carroll, Cline}. For a very recent and exhaustive review, we refer to the work of Wang, Abdalla, Atrio-Barandela and Pav\'{o}n (\cite{Wang:2016lxa}).

In the present work, we consider a dark energy fluid, with constant equation of state, which interacts with pressureless dark matter through a non-gravitational interaction.
Other components namely baryons and radiation are not considered for their minimal contribution to the total energy budget of the universe in compared to the combined dark sector (almost 96\% of the total energy density), see \cite{Planck2015}. The interaction term $Q$ is assumed to be a linear function of the energy densities of the dark components and their first order derivatives. The space-time has been considered to be characterized by a spatially flat Friedmann-Lema\^{i}tre-Robertson-Walker (FLRW) line element. With such linear interaction, the second order equation for the density is analytically solved. We analyze three interacting scenarios when the dark energy is of quintessential, vaccum energy and  phantom type, using the currently available observational data from  (i) Hubble parameter measurements and (ii) Supernovae type Ia from {\it joint light-curve analysis} (JLA). 
Data from baryon acoustic oscillations and cosmic microwave background radiation have not been used since the available data are not model independent. So the analysis presented in this work does not have any unnecessary bias towards the $\Lambda$CDM. As we have not considered the radiation and the baryonic matter in our energy budget, this exclusion should not lead to any inconsistency.  Apart from that we also discuss some other important  cosmological consequences of the model, for instance, the  thermodynamic implications, strength of the coupling through the estimation of coupling parameters, the direction of energy flow, and the geometric tests.  

 The manuscript is organized as the following. In section \ref{field-equations}, we describe the gravitational equations and introduce the generalized interaction in the dark sector.  Section \ref{technique} presents the analytic solutions of the background dynamics under consideration. Section \ref{observational-data} deals with a brief descriptions of the observational data employed in our analysis and the following section contains the observational constraints on the model parameters imposed by the data sets. Section \ref{aic-bic} deals with a possible model selection on the Akaike and Bayesian criteria. In section \ref{sec-coupling}, attempts have been made regarding the estimation of the coupling parameters. In section \ref{sec-geometrical tests}, the results given by the model are compared againts some geometrical tests. The section \ref{sec-thermo} deals with testing the models against the laws of thermodynamics, and in the last section \ref{discu}, we include a summary and some remarks on the results obtained.

\section{Dynamics of dark matter$-$dark energy interaction}
\label{field-equations}

We consider a spatially flat homogeneous and isotropic universe characterized by the line element,
\begin{equation}
{\rm d}s^2= -{\rm d}t^2 + a^2 (t) \Bigl[ {\rm d}r^2+  r^2 \left( {\rm d} \theta^2 + \sin^2 \theta \, {\rm d} \phi^2\right) \Bigr ],
\end{equation}
where $a(t)$ is the  {\it scale factor} of the universe. We consider that the matter sector of the universe comes from two barotropic fluids, namely the pressureless dark matter or cold dark matter (CDM) and dark energy (DE) represented by the following energy-momentum tensors,
$T_{\mu\nu}^{M}= \rho_m u_{\mu} u_{\nu}$ for CDM, and $T_{\mu\nu}^{D}= (p_d+\rho_d) u_{\mu} u_{\nu}+ p_d g_{\mu\nu}$ for DE, in which $(\rho_m, 0)$ are respectively the energy density and the pressure of the dust while $(\rho_d, p_d)$ represent the energy density and pressure of DE. Since the fluids are barotropic, hence, for CDM we have $p_m= w_m \rho_m$, and for DE, $p_d= w_d\, \rho_d$, where $w_m$ {\it (which is zero for CDM)} and $w_d$, are the equation of state (EoS) parameters for CDM and DE, respectively.  The Einstein's field equations read as

\begin{equation}
3 H^2 =  \rho_m + \rho_d ,
\label{friedmann1}
\end{equation}
and 
\begin{equation}
2 \dot{H} + 3 H^2  =-\, p_d ,\label{friedmann2}
\end{equation}
where an overhead dot represents the differentiation with respect to the cosmic time $t$ and $H= \dot{a}/a$ is the Hubble expansion rate. Throughout the paper we have fixed the units as, $c= 8 \pi G = 1$.  The conservation equation, $\nabla_{\mu} \left( T^{\mu\nu}_{m}+ T^{\mu\nu}_{d} \right)= 0$, for the composite fluid, can be written as

\begin{align}\label{conservation}
\dot{\rho}_m + \dot{\rho}_d + 3 H (\rho_m + \rho_d + p_d) & = 0,
\end{align}
which one can obtain by using equations (\ref{friedmann1}) and (\ref{friedmann2}).

Now, introducing the total energy density of the fluid as $\rho_t= \rho_m + \rho_d$ (Hence, the total pressure is identified as $p_t= p_m+p_d= p_d$, since the dark matter is pressureless),
the conservation equation (\ref{conservation}) can be written in  terms of $\rho_t$ as,

\begin{eqnarray}
&&\dot{\rho}_t+ 3 H (\rho_t+ w_d \rho_d)= 0 \nonumber\\ 
&&\Longleftrightarrow \dot{\rho}_t+ 3 H \left( 1+ \frac{w_d \rho_d}{\rho_t} \right) \rho_t = 0 \nonumber\\ 
&&\Longleftrightarrow \dot{\rho}_t+ 3 H \left( 1+ w_{tot} \right) \rho_t  = 0.\label{cons-total}
\end{eqnarray}
where $w_{tot}$ is the equation of state parameter for the composite fluid, i.e. $w_{tot}=p_t/\rho_t$. Now, it is readily seen that although the equation of state parameter for DE is constant but the composite fluid have a variable EoS parameter $w_{tot} =p_d/\rho_t$. Now, introducing the coincidence parameter $r= \rho_m/\rho_d$, total EoS parameter, $w_{tot}$, can be written as,

\begin{eqnarray}\label{eff-2fluid}
w_{tot}  = \frac{w_d \rho_d}{\rho_t} = \frac{w_d \rho_d}{\rho_m+ \rho_d}= \frac{w_d}{1+r}.
\end{eqnarray}
The net effective EoS parameter, $w_{tot}$, could describe an accelerating universe if $w_{tot}< -1/3$, which results in $w_d< -(1 + r)/3$.  Further, if one considers that $w_{tot}$ mimics the cosmological constant $\Lambda$ (i.e. $w_{tot} = -1$), then $w_d= -1-r$. If $w_d < - (1+r)$, the dark energy would cross the phantom divide, and the universe will end in a Big Rip.

Since the fluids are interacting via a non-gravitational interaction, the conservation equation  (\ref{conservation}) can be decoupled into the following two equations, 

\begin{align}
\dot{\rho}_m+ 3 H \rho_m &= Q,   \label{cons-DM}
\end{align}
and
\begin{align}
\dot{\rho}_d+ 3 H (1+ w_d) \rho_d &= -Q, \label{cons-DE}
\end{align}
in which $Q$ is the interaction function that determines the energy flow between the dark sectors, namely the DE and the CDM. In absence of the  interaction, i.e. when $Q= 0$, the dark components are independently conserved. The primary motivation behind introducing the interaction is simply not to ignore this possibility apriori and see if this interaction leads to any new features. For any arbitrary $Q$, one can write the evolution equations for CDM and DE in an implict way as
\begin{eqnarray}
\rho_m = \rho_{m0}\, a^{-3} + a^{-3}\,\int_{1}^{a} \left(\frac{Q}{a H} \right)a^3\, da,\label{new1}\\
\rho_d= \rho_{d0}\, a^{-3 (1+\omega_d)} - a^{-3 (1+\omega_d)} \int_{1}^{a} \left(\frac{Q}{aH}\right)a^{3(1+\omega_d)}\, da. \label{new2}
\end{eqnarray}
For $Q \propto H f(a)$, where $f(a)$ is any known analytic function, the equations (\ref{new1}) and (\ref{new2}) can be integrated. 
It is evident that the second terms in both equations (\ref{new1}) and (\ref{new2}) are the deviations from the standard evolution equations of CDM and DE, when they do not interact.\\

Since the interaction $Q$ is not really known, usually one starts with some ansatz for $Q$ and estimates the parameters in the ansatz with the help of  observational data. In the literature, extensive studies have been carried out with $Q \propto \rho_m$, $Q \propto \rho_d$, $Q \propto (\rho_m + \rho_d)$, and so on, where the proportionality constant, referred to as the coupling constant, determines the strength (from its magnitude) and direction (by its sign) of the interaction. In what follows, we consider a very general interaction of the form

\begin{align}\label{int}
Q & = 3\, H \, \lambda_m \, \rho_m+ 3\,H \, \lambda_d \, \rho_d + 3\, H \, \alpha_m \, \rho_m^\prime + 3\, H \, \alpha_d\, \rho_d^\prime,
\end{align}
where a prime denotes the differentiation with respect to $x= \ln \left(a/a_0 \right)$, $a_0$ being the scale factor at present epoch which we set to unity, and $\lambda_m$, $\lambda_d$, $\alpha_m$, $\alpha_d$ are the coupling constants. It is easy to see that by using the conservation equations (\ref{conservation})  (or eqns. (\ref{cons-DM}) and (\ref{cons-DE})), one arrives at

\begin{eqnarray}\label{sub-interaction}
Q = 3H (\lambda_m - 3\alpha_d)\rho_m+ 3 H  (\lambda_d- 3 \alpha_d (1+\omega_d)) \rho_d \nonumber\\+ 3H (\alpha_m -\alpha_d) \rho_m^\prime,
\end{eqnarray}
which directly includes the equation of state of dark energy\footnote{Note that any one amongst $\rho_m^{\prime}$ and $\rho_d^{\prime}$ can be used in terms of the other in view of equations (\ref{cons-DM}) and (\ref{cons-DE})}. Thus, the most general linear interaction of the form (in which the coefficients of $\rho_m$, $\rho_d$ and their derivatives are constants) is given as 
\begin{eqnarray}
Q&=& 3 \, H\, \lambda\, \rho_m+ 3\,H \,\mu\, \rho_d + 3\, H \, \alpha\, \rho_m^\prime,\label{interaction}
\end{eqnarray}
where ($\lambda$, $\alpha$, $\mu$) $\in$ ${\mathbb{R}}^3$ form the set of the coupling constants. In the following sections, the solution of the field equations  for this generalized interaction term and the observational constraints on the model have been discussed.

\section{Analytical solutions of the background evolution}
\label{technique}

It is  easy to find that from equation (\ref{cons-total}), one can readily write the energy densities of CDM and DE in terms of the total energy density ($\rho_t$) and its derivative $\rho_t^\prime$ in the following manner

\begin{eqnarray}\label{analytic}
\left(\rho_m,~~\rho_d \right)&=& \left(\frac{\rho_t^\prime+ 3 (1+\omega_d) \rho_t}{3 \omega_d},~~-~\frac{\rho_t^\prime+ 3 \rho_t}{3 \omega_d}\right).
\end{eqnarray}

Now, using the individual conservation equations (equations  (\ref{cons-DM}) and (\ref{cons-DE})),
we find that

\begin{align}
\rho_m ^\prime & = 3 \Bigl(  \bar{Q}- \rho_m  \Bigr),\label{sp1}\\
\rho_d^\prime & = -3 \Bigl(  \bar{Q}+ (1+ w_d)\, \rho_d  \Bigr)\label{sp2},
\end{align}
where $\bar{Q}= Q/ 3H$. Employing the above equations one can form the following second order differential equation in $\rho_t$,

\begin{align}\label{ode}
\rho_t^{\prime \prime}+ \Bigl[ 3+ 3 (1+ w_d)\Bigr]\, \rho_t^\prime +   9 (1+ w_d)\, \rho_t & = 9 \, w_d \, \bar{Q}. 
\end{align}

Using the interaction term, given in equation (\ref{interaction}), the differential equation (\ref{ode}) becomes,

\begin{eqnarray}\label{diffeqn-const-w}
 \rho^{\prime\prime}_t + 3  \Bigg[1+ \frac{(1+ \mu-\lambda)+ (1-3\alpha) w_d}{(1- 3 \alpha)}\Bigg]\,\rho^{\prime}_t \nonumber\\+ 9  \Bigg[\frac{(1+ w_d)(1-\lambda)+ \mu}{(1- 3 \alpha)}\Bigg]\,\rho_t & = 0,
\end{eqnarray}
which is a linear homogeneou second order differential equation with constant coefficients. The solution of the differential equation can be written if we know the roots of the auxiliary equation

\begin{eqnarray}\label{roots}
m^2+ 3\,\Bigg[1+ \frac{(1+ \mu-\lambda)+ (1-3\alpha) w_d}{1- 3 \alpha}\Bigg] m \nonumber\\+ 9\,\Bigg[\frac{(1+ w_d)(1-\lambda)+ \mu}{1- 3 \alpha} \Bigg] & = 0.
\end{eqnarray}

Thus, the total energy density takes the form

\begin{align}\label{unequal}
\rho_t & = \rho_1 (1+z)^{r_1}+ \rho_2 (1+z)^{r_2},
\end{align}
where $\rho_1$ and $\rho_2$ are the integration constants and both are positive ($\rho_1 > 0$, $\rho_2 > 0$)   \footnote{It should be noted that if one of the integration constants  is negative then at some epoch $\rho_t$ vanishes, which is unphysical, hence both the constants must be positive. Later we will indicate the physical meaning of the constant of integrations.}; $r_1$, $r_2$ are the roots of the auxiliary equation (\ref{roots}) and $z$ is the redshift. The expressions of $r_1$ and $r_2$ are given as,
\begin{eqnarray}
r_1   = \frac{3}{2} \Bigg[\Bigg(1+ \frac{(1+ \mu-\lambda)+ (1-3\alpha) w_d}{(1- 3 \alpha)}\Bigg) + \sqrt{X}\,\,\,\Bigg],\\
r_2   = \frac{3}{2} \Bigg[\Bigg(1+ \frac{(1+ \mu-\lambda)+ (1-3\alpha) w_d}{(1- 3 \alpha)}\Bigg)- \sqrt{X}\,\,\,\Bigg],
\end{eqnarray}
in which 
\begin{eqnarray}
X = \Bigg(1+ \frac{(1+ \mu-\lambda)+ (1-3\alpha) w_d}{(1- 3 \alpha)}\Bigg)^2\nonumber\\- 4 \Bigg(\frac{(1+ w_d)(1-\lambda)+ \mu}{(1- 3 \alpha)}\Bigg),
\end{eqnarray}
and this $X$ should be positive so that $r_1$, $r_2$ are real. This is quite a general case of interacting models. In fact, (i) for $\alpha= 0$, we get the interacting dynamics as discussed in \cite{Pan:2012ki,Pan:2016ngu}, (ii) $\alpha= \mu =0$ gives the interaction $Q= 3 H \lambda \rho_m$ \cite{He1}, (iii) for $\alpha= \lambda =0$, one realizes the dynamics for $Q= 3 H \mu \rho_d$ \cite{He1}. Therefore, the present setting covers a large class of interacting dynamics. Now, plugging (\ref{unequal}) into (\ref{analytic}), the evolution of CDM and DE can be expressed as,

\begin{eqnarray}
\rho_m  = \left(\frac{\rho_1\, \left(  3 (1+ w_d) -r_1 \right)}{3 w_d} \right) (1+z)^{r_1}\nonumber \\+ \left(\frac{\rho_2\, \left(  3 (1+ w_d) -r_2\right)}{3 w_d}\right) (1+z)^{r_2},\label{rhom}
\end{eqnarray}
\begin{eqnarray}
\rho_d  = \frac{\rho_1\, \left( r_1 -3  \right)}{3 w_d} (1+z)^{r_1}+  \frac{\rho_2\, \left(   r_2 -3  \right)}{3 w_d} (1+z)^{r_2}\label{rhod}.
\end{eqnarray}

Now, we introduce the density parameters as $(\Omega_{1}, \Omega_{2})= (\rho_1/\rho_c, \,\, \rho_2/\rho_c$) (where $\rho_c= 3H_0^2$). These parameters will have the expressions,

\begin{align}
\Omega_{m0} & = \Omega_{1}\,\left( 1+ \frac{3-r_1}{3\, w_d} \right )+ \Omega_{2}\,\left( 1+ \frac{3-r_2}{3\, w_d} \right ),\\
\Omega_{d0} & = \,\Omega_{1}\,\left( \frac{r_1-3}{3\, w_d} \right )+ \Omega_{2}\,\left(\frac{r_2-3}{3\, w_d} \right ),
\end{align}
where $\Omega_{m0}$, $\Omega_{d0}$, are the usual density parameters for matter and dark energy and for a spatially flat geometry at the present epoch (i.e. $z=0$) and are connected as 
$\Omega_{m0}+ \Omega_{d0}= \Omega_{1}+ \Omega_{2}= 1$, where

\begin{align}
\Omega_{1} & = \frac{3 (1+ w_d \, \Omega_{d0})-r_2}{r_1-r_2}~,
\end{align}
and
\begin{align}
 \Omega_{2} & = \frac{r_1 - 3 (1+ w_d \, \Omega_{d0})}{r_1 - r_2}.
\end{align}

Now, the equation (\ref{unequal}) can be written as
\begin{eqnarray}\label{hubble-expression}
\left(\frac{H}{H_0}\right)^2 = \Omega_{1}\, (1+z)^{r_1}+ \left(1- \Omega_{1}\right)\, (1+z)^{r_2}~.
\label{hubble}
\end{eqnarray}

Thus, for the interaction as in (\ref{interaction}), the evolution is actually analytically solved. The exponents $r_1$, $r_2$, contain the coupling parameters of the interaction, hence effectively they signal the deviation from the standard cosmology in presence of the interaction in the dark sector. For $r_1= 3$, and $r_2 =0$ (given by $\alpha=\mu=\lambda=0$, and $w_{d}=-1$), one recovers the standard $\Lambda$CDM cosmology. Any deviation from these values can be considered to be an indicator of the interaction. Now, since the background solution is analytic, one can find the associated cosmological parameters in this model. The conservation of all components of the cosmic fluid taken together given by equation (\ref{cons-total}) can be written as

\begin{equation}\label{t}
w_{tot} \equiv -1-\frac{\dot{\rho}_t}{3H \rho_t}= -1- \frac{2\dot{H}}{3H^2}= -1+ \left(\frac{1+z}{3H^2}\right)\, \frac{d}{dz} (H^2).
\end{equation}

Introducing the deceleration parameter $q= -1-\dot{H}/H^2$, one may also write

\begin{equation}\label{deceleration}
q= -1 + \left(\frac{1+z}{2 H^2}\right)\, \frac{d}{dz} (H^2),
\end{equation}
and consequently, the redshift, $z_t$, at which the universe entered into the current accelerating phase from the decelerating one (i.e. the transition redshift given by $q=0$) is found to be

\begin{equation}\label{tran-redshift}
z_t = \Bigg[\left(\frac{r_2-2}{2-r_1}\right)\,\left(\frac{r_1-3(1+w_d \Omega_{d0})}{3(1+w_d \Omega_{d0})-r_2}\right)\Bigg]^{\frac{1}{r_1- r_2}} - 1~.
\end{equation}

We close this section with an observation. If the interaction (\ref{int}) or (\ref{interaction}) allows the second order derivatives of the energy densities of the dark sectors, then the background evolution as well as the evolution equations for the dark matter and dark energy can be analytically solved using equations (\ref{analytic}). In a similar fashion, if the interaction function allows up to $n$-th order derivatives of the energy densities of the dark sectors, so that the interaction function remains linear, then the differential eqation (\ref{ode}) may be integrated for some particular values of $n$ leading to the analytic solutions of the background evolution. For any linear interaction $Q$ given by  $Q = Q (\rho_t, \rho_t^{(1)}, \rho_t^{(2)}, ...\rho_t^{(n)})$, where $\rho^{(i)}$ stands for the $i$-th derivative of $\rho_t$ with respect to $x = \ln a$, the differential equation will look like $A_0 \rho_t^{(n)} + A_{1} \rho_t^{(n-1)}+ A_{2} \rho_t^{(n-2)}+...+ A_{n-1} \rho_t + A_n = 0$, where $A_i$'s are all constants. Thus, for different values of $n$, it is possible to obtain analytic solutions for $\rho_t$ and hence using the equations in (\ref{analytic}), the evolution equations for $\rho_m$ and $\rho_d$ can also be analytically found. This is an interesting result for the linear interactions between dark matter and dark energy having constant equations of state.

\section{Statistical method and the observational data utilized in the analysis}
\label{observational-data}

We focus on the late-time behaviour of the interacting models. So, to constrain the proposed interacting models, we take two independent low-redshift observational data sets, namely the  (i) Supernovae type Ia, (ii) Hubble parameter measurements and restrict the analyses in an interval $0 \leq z < 2.34$.

In order to figure out the observational constraints on the cosmological parameters, we perform the fittings of the models using the {\it emcee}, introduced by \cite{ForemanMackey:2012ig}, the python implementation of the ensemble sampler for Markov chain Monte Carlo (MCMC). The estimation of the parameters follows the maximization of the likelihood function ${\mathcal L}=\exp{(-\chi^2_{tot}/2)}$ with $\chi^2_{tot}$ as, 

\begin{equation}
\chi^2_{tot}=\sum_j\chi^2_j~,
\end{equation}
where the index `$j$' denotes an individual data set. In general, the $\chi^2$ function is defined by the following relation

\begin{equation}
\chi^2({\theta})=\sum_i\frac{[\eta_{obs}(z_i)-\eta_{th}(\{\theta\},z_i)]^2}{\sigma^2_i},
\end{equation}
where $\eta_{th}$ is the theoretical value of some observable quantity, which is a function of the model parameters $\{\theta\}$, and $\eta_{obs}$ is the corresponding observational estimation at redshift $z_i$.\\

Using this likelihood analysis we aim to constrain three different interacting dark energy scenarios where the dark energy is either quintessence (i.e. $w_d>-1$), or the cosmological constant ($w_d=-1$), or phantom ($w_d<-1$). In order to facilitate the statistical analysis, specific values of $w_d$ will be chosen. \\

In a Bayesian statistical analysis, the prime idea is to figure out the posterior probability distribution of the parameter where the $posterior\sim prior\times likelihood$. In case of a uniform prior, the posterior distribution is proportional to the likelihood function. In what follows, a constant prior will be assumed for the parameter values. \\

A brief discussions about the observational data sets are presented in the following subsections.

\subsection{Supernoave Type Ia}
\label{sn-data}

Supernovae type Ia are the first observational data which indicated the alleged accelerated expansion of the universe and hence the existence of the dark energy. Here, we measure the distance modulus $\mu(z)$, of any Supernova of Type Ia located at redshift $z$, which is the difference between its apparent magnitude ($m_B$) and its absolute magnitude ($M_B$) of the B-band of the observed spectrum. It is defined by
\begin{equation}
\mu(z)=5\log_{10}{\Bigg(\frac{d_L(z)}{1Mpc}\Bigg)}+25,
\end{equation}
where the $d_L(z)$, the luminosity distance of  a particular supernoave of type Ia at redshift $z$, is defined in a spatially flat FLRW universe as
\begin{equation}
d_L(z)=(1+z)\int_0^z\frac{dz'}{H(z')}.
\end{equation}
In the current work, we use the 31 binned distance modulus data sample of the recent joint light curve analysis (JLA) \cite{betoule}, and to encounter the correlation between different bins, we adopt the prescription given in \cite{farooqmaniaratra}.

\subsection{Hubble parameter measurements}
\label{ohd}

The observational Hubble parameter data (OHD) is one of the most robust probes to analyze the dark energy models for its model independent nature. Here, 29 measurements of the Hubble parameter from different surveys have been used. We adopt the values of $H (z)$  which have been estimated from the measurements of differential of redshift $z$ with  respect to the cosmic time $t$ as follows
\begin{equation}
H(z)=-\frac{1}{(1+z)}\frac{dz}{dt}.
\label{Hz}
\end{equation}
In general there are several ways to measure the Hubble parameter values at different redshifts. Here, we use the Hubble parameter measurements using
the differential age of galaxies estimator of $\frac{dz}{dt}$  by Simon {et al} \cite{simon}; the red-enveloped galaxies by Stern {\it et al} \cite{stern}; the measurement of Hubble parameter values at low redshift using the differential age method along with Sloan Digital Sky Survey (SDSS) data by Zhang {\it et al} \cite{zhang}; the measurements from the old and passively evolving galaxies (known as cosmic chronometers) by Moresco {\it et al} \cite{moresco}, and finally, the measurement at a high redshift $z=2.34$, by Delubac {\it et al} \cite{delubac}. The measurement of $H_0$ from the last Planck results \cite{Planck2015} has also been used in the analysis.

\begin{table*}
	\begin{center}
		\caption{Results of the statistical analysis for the interacting quintessence dark energy model. The reduced $\chi^2$, i.e. $\chi^2_{min}/d.o.f.$ and the values of the model parameters (at 1$\sigma$ error bar) obtained for different combinations of the data sets have been presented.}
		\label{tableResult1}
		\resizebox{0.5\textwidth}{!}{  
			\begin{tabular}{ c |c |c c c } 
				\hline
				\hline
				& $\chi^2_{min}/d.o.f.$ & $\Omega_{d0}$  & $~~r_1$ & $~~r_2$ \\ 
				\hline
				
				OHD & 0.514 & $0.756^{+0.102}_{-0.102}$ & ~~$2.708^{+0.378}_{-0.246}$ & ~~$-0.763^{+0.969}_{-0.824}$\\ 
				
				SNe & 1.21 & $0.669^{+0.062}_{-0.081}$ & ~~$2.477^{+0.656}_{-0.356}$ & ~~$0.190^{+0.474}_{-0.721}$\\

				SNe$+$OHD & 0.834 & $0.717^{+0.044}_{-0.046}$ & ~~$2.787^{+0.290}_{-0.228}$ & ~~ $-0.227^{+0.422}_{-0.478}$\\ 
				\hline
				\hline
			\end{tabular}
		}
	\end{center}
	
\end{table*}
\begin{figure*}
		\includegraphics[width=0.75\textwidth]{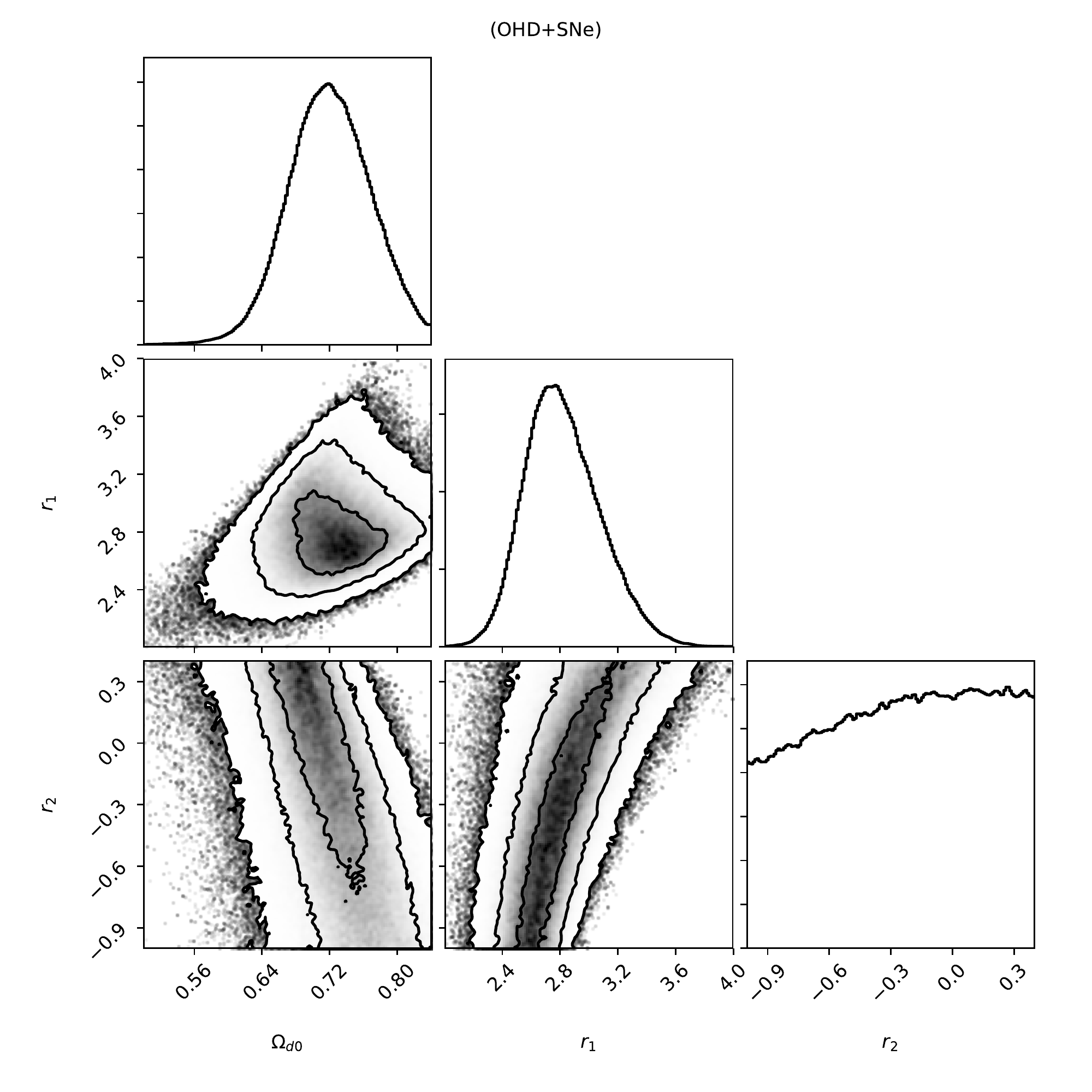}
	\caption{\textit{$1\sigma$ ($68.3\%$), $2\sigma$ ($95.4\%$) and $3\sigma$ ($99.7\%$) confidence level contour plots for the model parameters of the interacting quintessence dark energy, in particular for $w_{d}=- \, 0.98$, have been shown using the observational data OHD$+$SNe. Additionally, the figure also shows the one dimensional marginalised posteriors distributions for the parameters ($\Omega_{d0}$, $r_1$, $r_2$). }}
	\label{quintCont}
\end{figure*}
\begin{figure}
	\begin{center}
		\includegraphics[angle=0, width=0.4\textwidth]{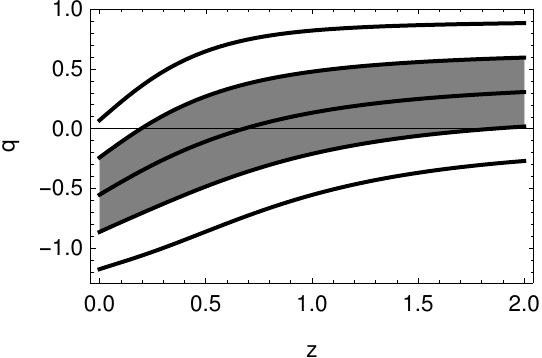}
		\includegraphics[angle=0, width=0.4\textwidth]{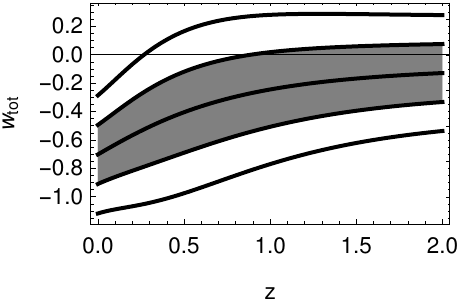}
	\end{center}
	\caption{\textit{The plots for deceleration parameter $q$ (upper panel) and the total equation of state $w_{tot}$ (lower panel) in the interacting scenario when dark energy is of quintessence type, have been shown using the observational data OHD$+$SNe. In both panels we have shown the 1$\sigma$ ($68.3\%$) and 2$\sigma$ ($95.4\%$) confidence regions around the best fit curve (the central dark line).}}
	\label{quintqzwtot}
\end{figure}

\section{Results of the Statistical analysis}
\label{results}

In the present work, the statistical analysis involves the estimation of the independent parameters  present in the expression of the Hubble parameter in eqn. (\ref{hubble}). The parameter $\Omega_1$ can be written in terms of $r_1$, $r_2$, $\Omega_{d0}$ and $w_d$. The dark energy density parameter $\Omega_{d0}$ and dark energy equation of state parameter $w_d$ are coupled to each other in this case and thus can not be estimated independently. In the present analysis, three different values of $w_d$ have been assumed, which correspond to the three different dark energy scenario, namely the {\it quintessence}, the  {\it vacuum energy density} and the {\it phantom} dark energy.

\subsection{Interacting quintessence ($w_d>-1$)}
\label{sec-int-quintessence}

For a quintessence model, the value of the equation of state parameter  lies in the range $\left(-1,  -1/3  \right)$. In the present analysis, we have fixed $w_d= -0.98$. This typical value of $w_d$ has been motivated from the recent observations \cite{Planck2015} that the dark energy equation of state is very close to the cosmological constant boundary `$w_d = -1$' \footnote{From the latest estimation by Planck \cite{Planck2015} the equation of state for dark energy is 
constrained as $w_d = -1.006 \pm 0.045$.}.
The results of the estimation of the free parameters of the model from the observational data sets, namely (i) SNe, (ii) OHD, (iii) OHD and SNe combine, are presented in Table \ref{tableResult1}. The contour plots of various quantities as well as their likelihood analysis have been shown in Fig. \ref{quintCont}, obtained in the combined analysis with OHD$+$SNe. Additionally, in Fig. \ref{quintqzwtot}, using the same combined analysis OHD$+$SNe, we have shown the transition of the deceleration parameter (upper panel of Fig. \ref{quintqzwtot}) and the qualitative evolution of the total equation of state parameter  (lower panel of Fig. \ref{quintqzwtot}).  \\

From the analyses summarized in Table \ref{tableResult1}, 
one may notice that the 
observational data OHD$+$SNe significantly decrease the error bars on the parameters ($\Omega_{d0}$, $r_1$, $r_2$) compared to the separate analysis performed either with SNe or OHD. From Table \ref{tableResult1}, one can see that data from OHD estimate the mean value of $r_2$ to be negative while the SNe data predict the mean value of $r_2$ to be positive. The mean value of $r_2$ is negative for the combined analysis OHD$+$SNe. From the analyses one may notice that the positive values of $r_2$ are still allowed within $1\sigma$ confidence level which is also clearly seen from Fig. \ref{quintCont}. The value of $r_1$ is strictly positive within the $1\sigma$ confidence level for all the data sets used in the analysis. Thus, effectively, one can see that within $1\sigma$ confidence level, $r_1 > 0$ while $r_2$ is allowed to have both positive and negative values. This is indeed a saviour of the model {since} a negative definite value of $r_2$ would mean some component of matter density increases with the evolution of the universe!\\

Furthermore, from the constraints on $r_1$, $r_2$ as shown in Table \ref{tableResult1}, the effects of interaction is visible. One can see that all combinations of the observational data suggest that the values $r_1 \neq 3$, and $r_2 \neq 0$ are quite a possibility, so an interaction is not at all ruled out. However, it is interesting to note that all analyses do allow the $\Lambda$CDM limit ($r_1 = 3$, $r_2 = 0$) in the $1\sigma$ confidence-level. Additionally, from the estimations of $r_1$ and $r_2$, it is also clear that the combined data sets of  OHD and SNe brings the model closer to the $\Lambda$CDM limit compared to the individual data sets. 
Further, from the evolution of the deceleration parameter (upper panel of Fig. \ref{quintqzwtot}), it is clear that the transition from decelerating phase to the current accelerating phase happened for $z \in (0.6, 0.9)$ which is in good agreement with several observational results. Additionally, from the evolution of $w_{tot}$ (lower panel of Fig. \ref{quintqzwtot}), one can see that crossing the phantom devide is allowed only in the $2\sigma$ confidence level. 
The plots in Fig. \ref{quintCont} also indicate that ($r_1 , {\Omega}_{d0}$) as well as ($r_1 , r_2$) are positively correlated whereas ($r_2 ,{\Omega}_{d0}$) are negatively correlated.

\begin{table*}
	\begin{center}
		\caption{Results of the statistical analysis for the interacting vacuum energy density model. The reduced $\chi^2$, i.e. $\chi^2_{min}/d.o.f.$ and the values of the mode parameters (at 1$\sigma$ error bar) obtained for different combinations of the data sets have been presented.}
		\label{tableResult2}
		\resizebox{0.5\textwidth}{!}{  
			\begin{tabular}{ c |c |c c c } 
				\hline
				\hline
				Data   & $\chi^2_{min}/d.o.f.$ & $\Omega_{d0}$  & $~~r_1$ & $~~r_2$ \\ 
				\hline
				
				OHD & 0.526 & $0.744^{+0.102}_{-0.102}$ & ~~$2.713^{+0.363}_{-0.248}$ & ~~$-0.772^{+0.958}_{-0.819}$\\ 
				
				SNe & 1.211 & $0.656^{+0.064}_{-0.091}$ & ~~$2.449^{+0.633}_{-0.336}$ & ~~$0.160^{+0.425}_{-0.703}$\\ 
				
				SNe$+$OHD & 0.835 & $0.704^{+0.047}_{-0.047}$ & ~~$2.790^{+0.288}_{-0.288}$ & ~~$-0.241^{+0.436}_{-0.485}$\\ 
				\hline
				\hline
			\end{tabular}
		}
	\end{center}
	
\end{table*}
\begin{figure*}
		\includegraphics[width=0.75\textwidth]{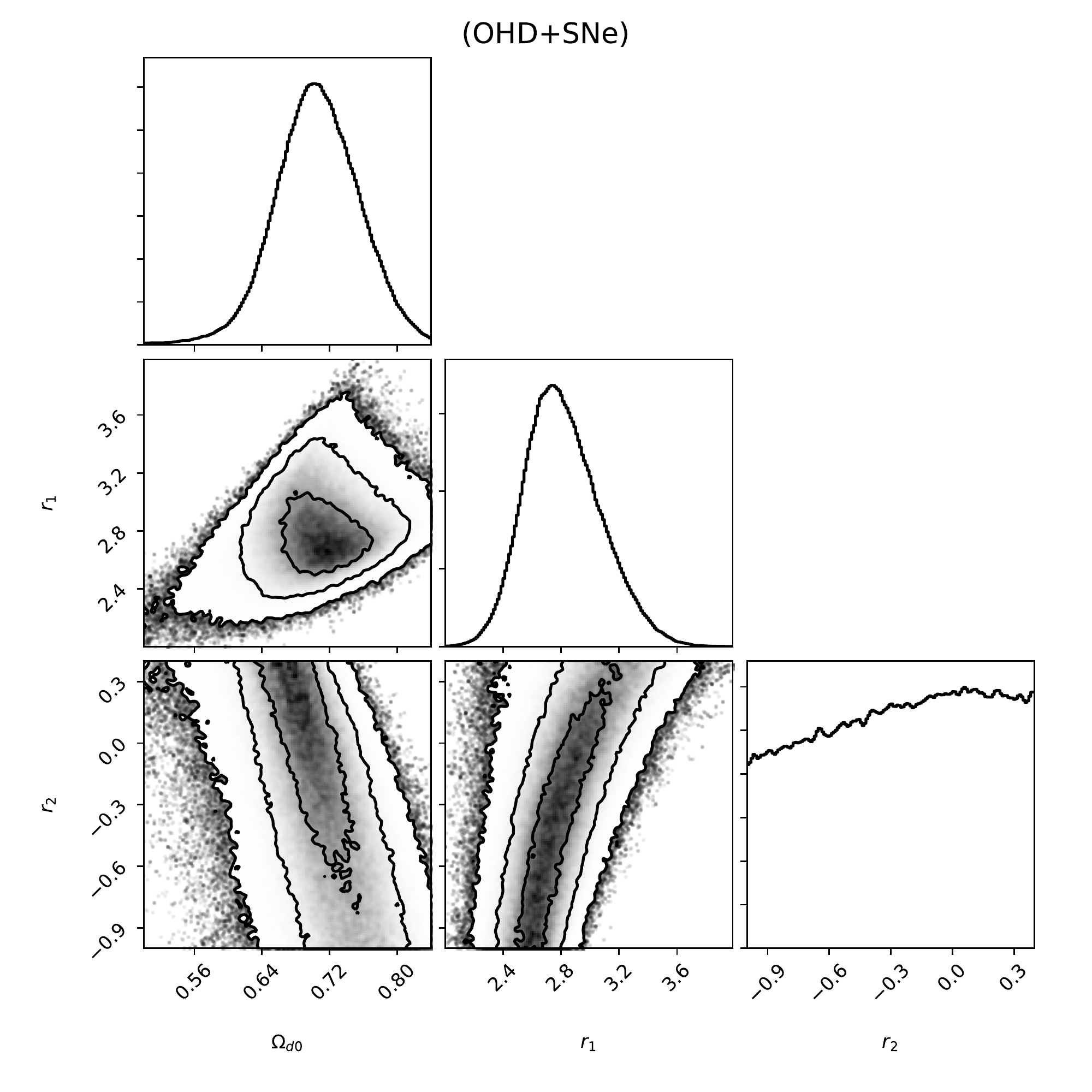}
	\caption{\textit{$1\sigma$ ($68.3\%$), $2\sigma$ ($95.4\%$) and $3\sigma$ ($99.7\%$) confidence level contour plots for the model parameters of the interacting cosmological constant (i.e. $w_d= -\,1$) using the observational data OHD$+$SNe. Additionally, the figure also displays the one dimensional marginalised posteriors distributions of the parameters for the parameters ($\Omega_{d0}$, $r_1$, $r_2$). }}
	\label{cosCont}
\end{figure*}
\begin{figure}
	\begin{center}
		\includegraphics[angle=0, width=0.4\textwidth]{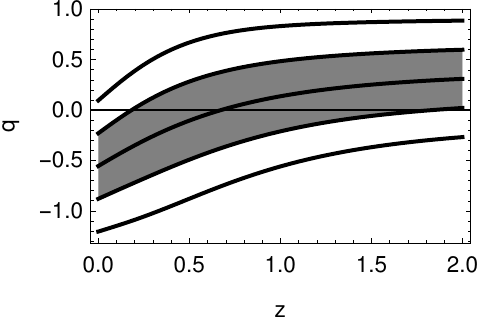}
		\includegraphics[angle=0, width=0.4\textwidth]{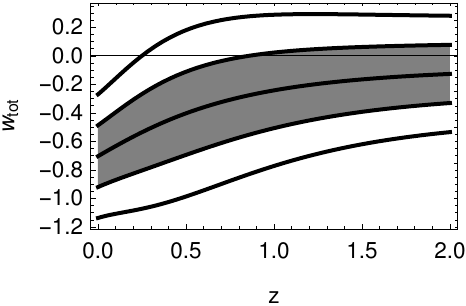}
	\end{center}
	\caption{\textit{The plots for the deceleration parameter $q$ (upper panel) and the total equation of state $w_{tot}$ (lower panel) in the interacting scenario when dark energy is the cosmological constant itself, have been shown using the observational data OHD$+$SNe. In both panels we have shown the 1$\sigma$ ($68.3\%$) and 2$\sigma$ ($95.4\%$) confidence regions around the best fit curve (the central dark line). }}
	\label{cosqzwtot}
\end{figure}


\subsection{Interacting vacuum energy density ($w_d=-1$)}
\label{sec-int-Lambda}

For vacuum energy density, $w_d = -1$, and here we assume that the vacuum dark energy is interacting with dust matter, the normal conservation equation, $\dot{\rho}+3H(\rho +p)=0$, does not hold, and ${\rho}_{vac}$ is no longer a constant. So, ${\rho}_{vac}$ cannot be identified with $\Lambda$. The results of free parameters of the model obtained from the analysis with different observational data sets, namely (i) OHD, (ii) SNe, and (iii) OHD$+$SNe are presented in the Table \ref{tableResult2}. The contour plots of various quantities as well as their likelihood analysis have been shown in Fig. \ref{cosCont} for combined data of OHD$+$SNe. Using the same combined data sets, OHD$+$SNe, we have shown the transition of the deceleration parameter (upper panel of Fig. \ref{cosqzwtot}) and the evolution of the total equation of state parameter (lower panel of Fig. \ref{cosqzwtot}).  \\

The analysis with OHD$+$SNe data sets for this interacting scenario significantly improves the error bars in the cosmological parameters ($\Omega_{d0}$, $r_1$, $r_2$) in a similar fashion observed in the previous model with $w_d > -1$. Moreover, we find that the behaviour of $r_2$ is exactly same as that observed in the interacting quintessence model and $r_1$ is strictly positive (within $1\sigma$ confidence level) for all the observational data sets employed in the analysis.  \\

From Table \ref{tableResult2}, we see that the mean value of $r_1 \neq 3$ and that of $r_2 \neq 0$ as suggested by all combined analyses. That means this interaction scenario is distinct from $\Lambda$CDM model which is characterized by $r_1 =3$ and $r_2 =0$. However, one may also note that the value of $r_1 = 3$ and $r_2 = 0$ is indeed allowed in the $1\sigma$ confidence-level for all the analyses performed in this work. From the upper panel of Fig. \ref{cosqzwtot}, it is clear that the transition from decelerating phase to the current accelerating phase happened for $z \in (0.6, 0.8)$ which again is in agreement with several observational results. From the evolution of $w_{tot}$ depicted in the lower panel of Fig. \ref{cosqzwtot} one finds a similar pattern as observed in the interacting quintessence model, that means, the phantom character of the composite fluid is still allowed (in $2\sigma$). 
It is easy to see that $r_1$ and ${\Omega}_{d0}$ are positively correlated while $r_2$, $\Omega_{d0}$ are negatively correlated but $r_1$ and $r_2$ are positively correlated amongst themselves. 

\begin{table*}
	\begin{center}
		\caption{Results of the statistical analysis for an interacting phantom dark energy model. The reduced $\chi^2$, i.e. $\chi^2_{min}/d.o.f.$ and the values of the mode parameters (at 1$\sigma$ error bar) obtained for different combinations of the data sets have been presented.}
		\label{tableResult3}
		\resizebox{0.5\textwidth}{!}{  
			\begin{tabular}{ c |c |c c c } 
				\hline
				\hline
				Data  & $\chi^2_{min}/d.o.f.$ & $\Omega_{d0}$  & $~~r_1$ & $~~r_2$ \\ 
				\hline
		
				OHD & 0.522 & $0.728^{+0.102}_{-0.099}$ & ~~$2.714^{+0.384}_{-0.248}$ & ~~$-0.775^{+0.982}_{-0.823}$\\ 
				
				SNe & 1.210 & $0.602^{+0.090}_{-0.083}$ & ~~$2.422^{+0.583}_{-0.313}$ & ~~$0.395^{+0.288}_{-0.778}$\\ 
				
				SNe$+$OHD  & 0.833 & $0.691^{+0.048}_{-0.049}$  & ~~$2.788^{+0.287}_{-0.230}$ & ~~$-0.236^{+0.436}_{-0.490}$ \\ 
				\hline
				\hline
			\end{tabular}
		}
	\end{center}
	
\end{table*}
\begin{figure*}
		\includegraphics[width=0.75\textwidth]{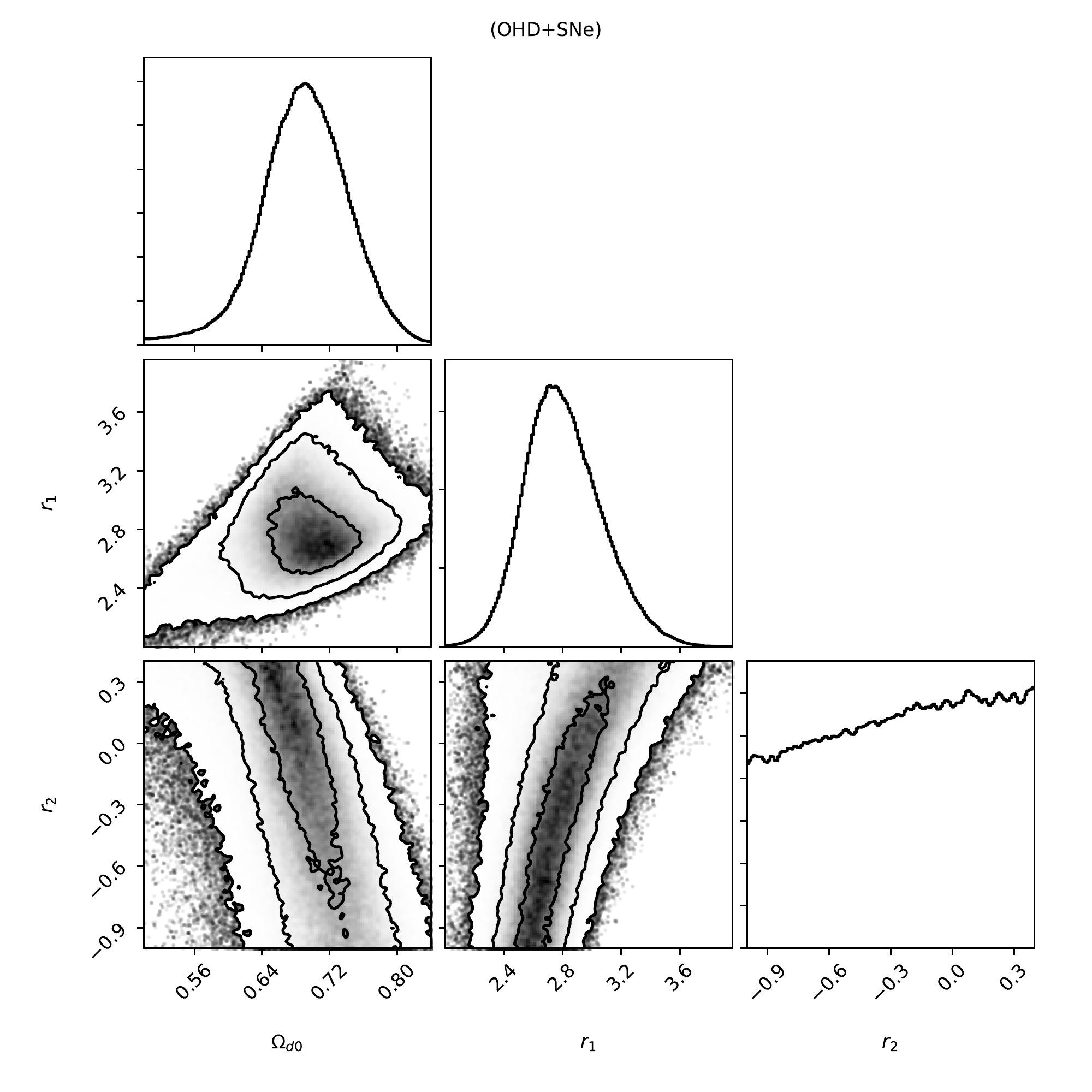}
	\caption{\textit{$1\sigma$ ($68.3\%$), $2\sigma$ ($95.4\%$) and $3\sigma$ ($99.7\%$) confidence level contour plots for the model parameters of the present interacting phantom dark energy, in particular for $w_{d}=-1.02$, have been shown using the observational data OHD$+$SNe. Additionally, the figure also shows the one dimensional marginalised posteriors distributions for the parameters ($\Omega_{d0}$, $r_1$, $r_2$). }}
	\label{phantCont}
\end{figure*}
\begin{figure}
	\begin{center}
		\includegraphics[angle=0, width=0.4\textwidth]{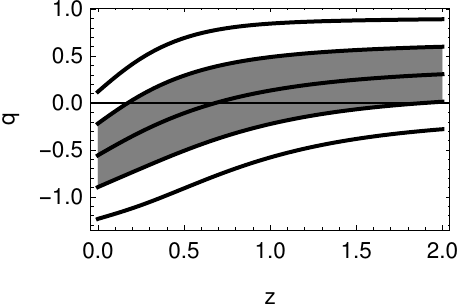}
		\includegraphics[angle=0, width=0.4\textwidth]{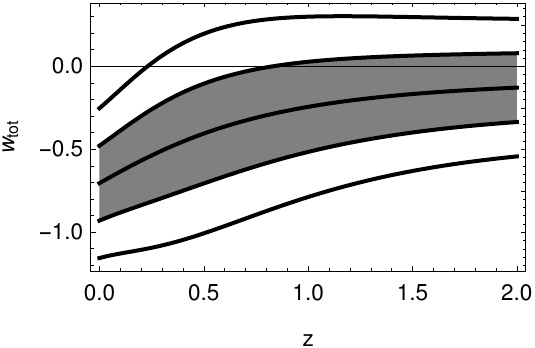}
	\end{center}
	\caption{\textit{The plots for deceleration parameter $q$ (upper panel) and the total equation of state $w_{tot}$ (lower panel) in the interacting scenario when dark energy is of phantom type, have been shown using the observational data OHD$+$SNe. In both panels we have shown the 1$\sigma$ ($68.3\%$) and 2$\sigma$ ($95.4\%$) confidence regions around the best fit curve (the central dark line).}}
	\label{phantqzwtot}
\end{figure}
\subsection{Interacting phantom dark energy ($w_d<-1$)}
\label{sec-int-phantom}

If the dark energy has a phantom nature, i.e., $w_d  < -1$, the volume and the rate of increase of the volume of the universe can blow up at a finite future depending on the total equation of state parameter. For an analysis of such a dark energy, interacting with the cold dark matter, we have fixed $w_d= -1.02$. This is phantom in nature, but close to $-1$ nonetheless and thus consistent with the observational requirement that it should be close to $-1$. The values of the free parameters of the model, extracted from different observational data sets, namely (i) OHD, (ii) SNe, (iii) OHD$+$SNe, are presented in the Table \ref{tableResult3}. The contour plots of various quantities as well as their likelihood analysis have been shown in Fig. \ref{phantCont}, obtained from the combined analysis with OHD$+$SNe. Also, in Fig. \ref{phantqzwtot}, we have shown the transition of the deceleration parameter (upper panel of Fig. \ref{phantqzwtot}) and the qualitative evolution of the total equation of state parameter (lower panel of Fig. \ref{phantqzwtot}) for the combined analysis
OHD$+$SNe. \\

The best constraints on the model parameters are obtained for the combined analysis OHD$+$SNe and the behaviour of $r_2$ follow the similar pattern as already observed in our previous two models. In addition, similar to the other two interaction models, here too, within $1\sigma$ confidence level, $r_1$ is strictly positive for all the observational data. \\

The values on $r_1$ and $r_2$ follow similar trend as observed in the previous two interacting scenarios, that means, $r_1 \neq 3$ and $r_2 \neq 0$ are suggested by all the analysis given in Table \ref{tableResult3} which shows that a  deviation from the $\Lambda$CDM cosmology is indeed allowed. However, the $\Lambda$CDM limit is still very much a possibility in the $1\sigma$ confidence-level. The departure from the $\Lambda$CDM cosmology is actually minimal for the combined analysis OHD$+$SNe. Further, from the evolution of the deceleration parameter (upper panel of Fig. \ref{phantqzwtot}), it is clear that the transition from decelerating phase to the current accelerating phase happened for $z \in (0.6, 0.8)$ which is in agreement with observational results. Moreover, it is important to note that the behaviour of the total equation of state parameter  $w_{tot}$ is still very similar to the other two cases despite the DE is assumed to be phantom. The phantom character for $w_{tot}$ is allowed only in a $2\sigma$ confidence-level. Finally, we note that the correlations between several combinations of the parameters $r_1$, $r_2$, $\Omega_{d0}$ show a similar pattern as observed in the interacting quintessence and interacting vacuum models.

\begin{figure*}
	\begin{center}
		\includegraphics[width=0.32\textwidth]{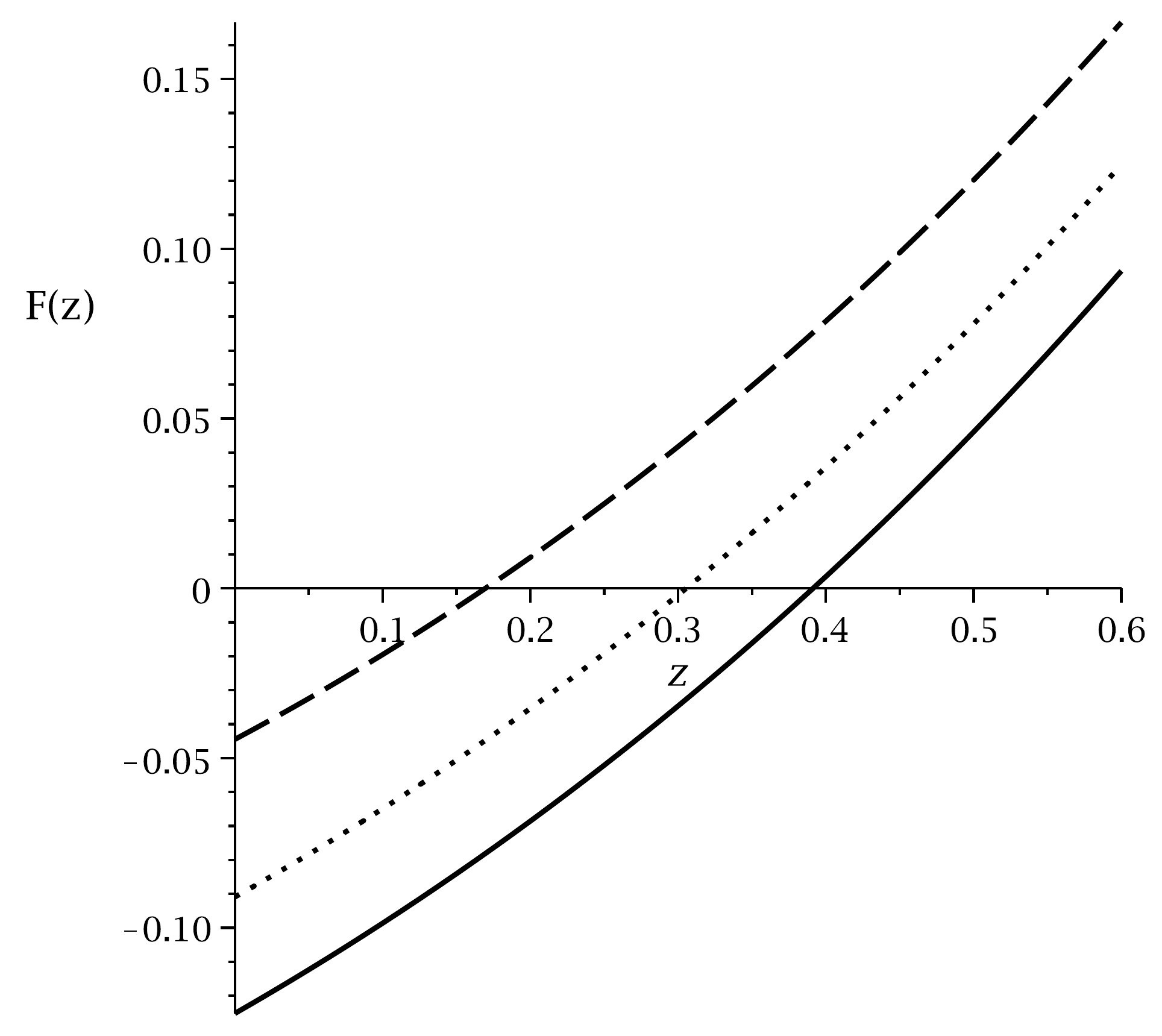}
		\includegraphics[width=0.32\textwidth]{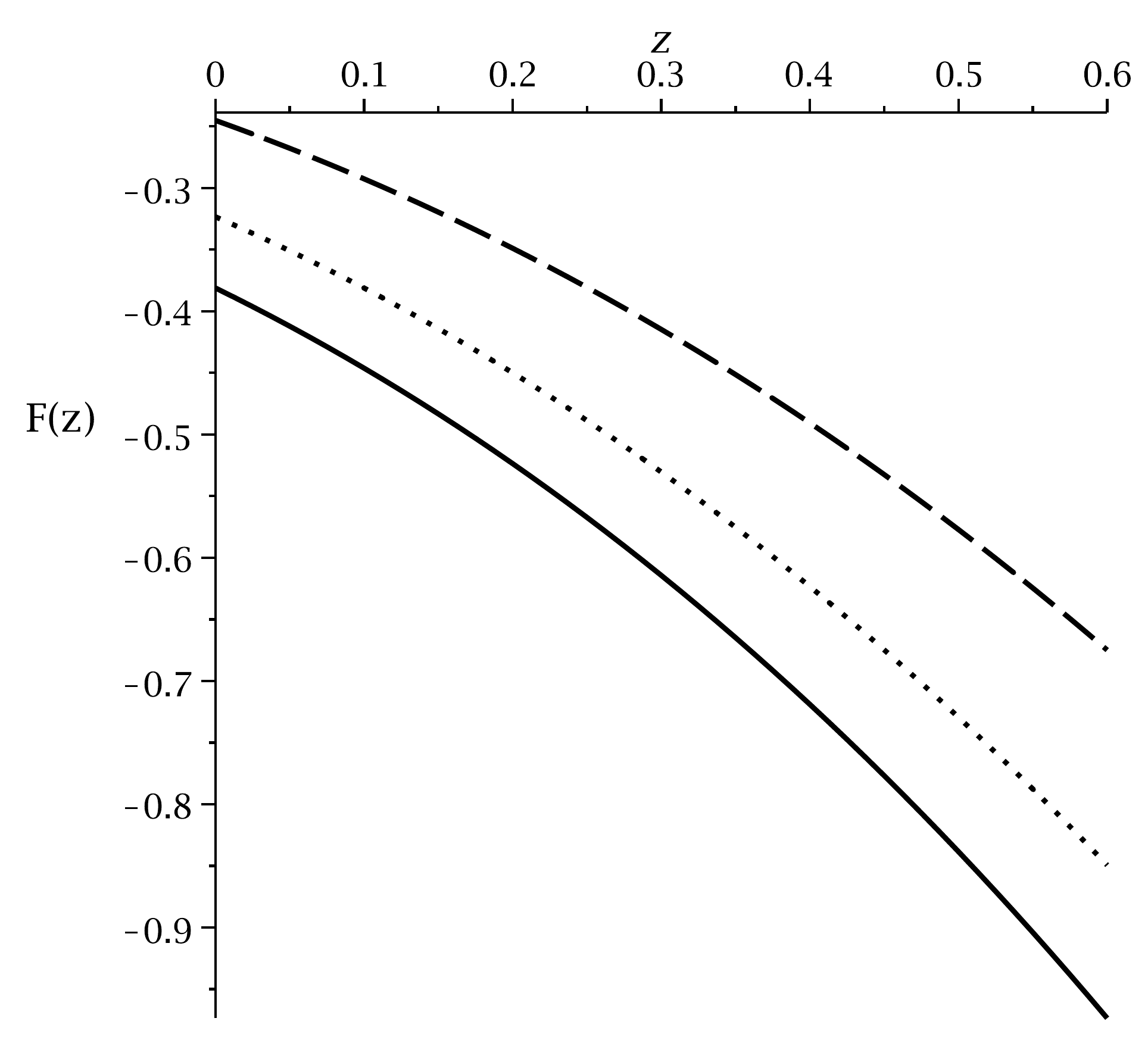}
		\includegraphics[width=0.30\textwidth]{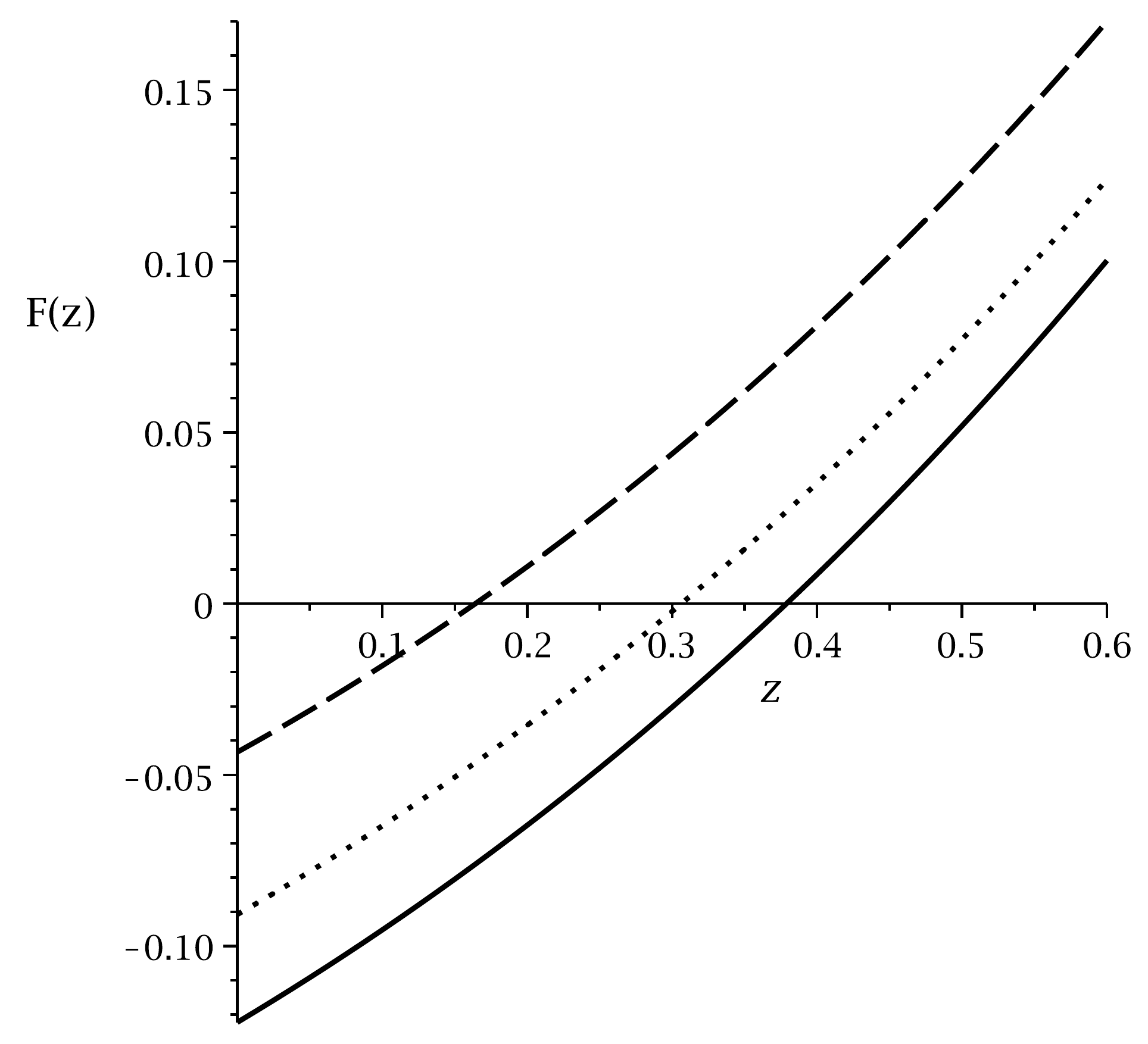}
	\end{center}
	\caption{The left, middle and the right panels of this figure respectively describe the behavior of the interaction function (\ref{int-A}), (\ref{int-B}) and (\ref{int-C}). In each plot the solid, dot and dashed lines respectively stand for the intracting quintessence, interacting vacuum and the interacting phantom models for the observational data OHD$+$SNe.}
	\label{figure-intABC}
\end{figure*}

\section{Information criteria and model selection}
\label{aic-bic}

Amongst several competing cosmological models a hierachy of the preferred models supported by the observational data is indeed required in the absence of any definite knowledge about the actual model. The commonly known  methods are the Akaike Infor mation Criterion (\textit{AIC}) \cite{aic} and the Bayesian Information Criterion (\textit{BIC}) \cite{bic}. The information criteria tells us which model is better. This model selection can also help us to rule out which models can be excluded. For a cosmological model with $d$ of degrees of freedom in which $N$ number of data points have been used to fit the model, AIC and BIC are respectively defined as

\begin{equation}
AIC = -2 \, \ln \mathcal{L}_{max}+ 2\, d,
\end{equation}

\begin{equation}
BIC = -2 \, \ln \mathcal{L}_{max}+ d\, \ln N,
\end{equation}
where $\mathcal{L}_{max}$ is the maximum likelihood obtained for the cosmological model. However, there is another method which is the corrected version of $AIC$, denoted by $AIC_c$ and defined as \cite{aicc}

\begin{equation}
AIC_c = AIC+ \frac{2\,d \,(d+1)}{N-d-1}~,
\end{equation}
which clearly shows that for $N \gg d$, $AIC_c$ is almost equal to the $AIC$. However, for small $N$, the difference between them is significant and this corrected version works better than the $AIC$ \cite{Liddle}.  It requires a `reference model' for comparison, and $\Lambda$CDM could be the best reference model and we choose that for the purpose. The rest of the analysis is simple. Let us denote any cosmological model by $M$, and then we calculate the difference $\Delta X= X_{M}- X_{\Lambda CDM}$ (where $X_{M} = AIC$, $BIC$, or $AIC_{c}$). Now, according to Jeffrey's scale, the significance of any cosmological model in comparison with the base model is measured from the difference $\Delta X$, where  $\Delta X > 5$, and $\Delta X > 10$ stand for strong and decisive evidence against the cosmological model under consideration \cite{Liddle}. 
Table \ref{tab-IC} summarizes the information criteria for different models with respect to the $\Lambda$CDM cosmological model. Our analysis shows that for all interacting models $\Delta X < 5$ (for $X= AIC,\, AIC_c$) for three different data sets, while $ 5< \Delta BIC < 10$, for the three data sets employed in the work. Although $AIC$ or $AIC_c$ criteria show that the interacting models do not have any strong evidence against them vis-a-vis the $\Lambda$CDM, but the BIC criterion tells us that the interacting models have strong evidence against them as compared to the $\Lambda$CDM but are not ruled out as comapred to the base $\Lambda$CDM model. 

\begin{table*}
	\begin{center}
	\caption{\label{tab-IC}\textit{Summary of the $\Delta AIC$, $\Delta BIC$, and $\Delta AIC_c$ values of the interacting dark energy models for the data sets   OHD, SNe, and OHD$+$SNe.}}
		\begin{tabular}{|r|c|c|c|c|c|c|c|c|c|c|}
			
			\hline \small
			\footnotesize Models & \multicolumn{3}{c|}{ \footnotesize OHD}&\multicolumn{3}{c|}{\footnotesize SNe} &\multicolumn{3}{c|}{\footnotesize OHD$+$SNe}\\
			\cline{2-3}
			\cline{4-5}
			\cline{5-6}
			\cline{6-7}
			\cline{7-8}
			\cline{8-10}
			&\footnotesize $\Delta AIC$ & \footnotesize 
			$\Delta BIC$ & \footnotesize $\Delta AIC_c$
			&\footnotesize $\Delta AIC$ & \footnotesize 
			$\Delta BIC$ & \footnotesize 
			$\Delta AIC_c$ & \footnotesize $\Delta AIC$ & \footnotesize $\Delta BIC$ & \footnotesize $\Delta AIC_c$\\
			\hline
			\hline
			\footnotesize Interacting quintessence &{\footnotesize $4.02$}&{\footnotesize $6.75$}&{\footnotesize $4.83$}&{\footnotesize $4.03$}&{\footnotesize $6.90$}&{\footnotesize $4.78$}&{\footnotesize $4.31$}&{\footnotesize $8.50$}&{\footnotesize $4.67$}\\
			\footnotesize Interacting vacuum &{\footnotesize $4.33$}&{\footnotesize $7.06$}&{\footnotesize $5.14$}&{\footnotesize $4.04$}&{\footnotesize $6.91$}&{\footnotesize $4.79$}&{\footnotesize $4.36$}&{\footnotesize $8.55$}&{\footnotesize $4.72$}\\
			\footnotesize Interacting phantom &{\footnotesize $4.22$}&{\footnotesize $6.95$}&{\footnotesize $5.03$}&{\footnotesize $4.01$}&{\footnotesize $6.88$}&{\footnotesize $4.76$}&{\footnotesize $4.28$}&{\footnotesize $8.47$}&{\footnotesize $4.64$}\\
			\hline
		\end{tabular}
	\end{center}
\end{table*}

\section{Estimations of the coupling parameters}
\label{sec-coupling}

We showed that for the interaction given by equation (\ref{interaction}) one can solve the evolution analytically, where we introduced two constant parameters $r_1 = r_1(\lambda, \mu, \alpha, w_d)$, $r_2= r_2 (\lambda, \mu, \alpha, w_d)$ which are in general determined by the constant coupling parameters as well as the dark energy equation of state. Now, in this work we  have considered three different interacting dark energy models characterized by three distinct numerical values of the dark energy equation of state, $w_d$. Nevertheless,
it is clear that if we use the estimates of $r_1$ and $r_2$ as given by our statistical analysis, all the three coupling parameters cannot be evaluated using two equations. One of them has to be chosen as an ansatz. In what follows, we shall discuss three simple special cases of the general interaction as given in equation (\ref{interaction}). \\

1. When $\alpha = 0$, the interaction (\ref{interaction}) becomes 

\begin{equation}\label{int-A}
Q= 3 H (\lambda \rho_m + \mu  \rho_d),
\end{equation} 
and consequently, one can express the coupling parameters as 

\begin{align}
\lambda &= \frac{1}{3 w_d}\, \left[r_1 + r_2 - \frac{r_1 r_2}{3} -3\right],\\
\mu &=  \frac{1}{3 w_d}\, \left[r_1 + r_2 - \frac{r_1 r_2}{3} -3\right] + \frac{r_1+r_2}{3}- 2 -w_d,
\end{align}
which clearly show that once $r_1$, $r_2$ are estimated, the coupling parameters can all be determined if $w_d$ is known. We have already chosen some typical values of $w_d$ for various distinct types of dark energy. For the joint analysis OHD$+$SNe, the coupling parameters can be estimated as (i) for interacting quintessence dark energy, the values of the parameters are, $\lambda=0.078^{+0.107}_{-0.102}$ and
$\mu=-0.089^{+0.135}_{-0.128}$; (ii) for interacting vacuum dark energy, the values are,  $\lambda=0.076^{+0.130}_{-0.100}$ and $\mu= -0.075^{+0.142}_{-0.127}$. Finally, (iii) for the interacting phantom dark energy, the values are,  $\lambda= 0.075^{+0.104}_{-0.098}$ and $\mu= -0.055^{+0.144}_{-0.135}$. \\

2. The case for $\lambda = 0$,  leads to the following interaction 

\begin{eqnarray}\label{int-B}
Q= 3 H (\mu \rho_d + \alpha \rho_m^\prime),
\end{eqnarray}
for which the coupling parameters become

\begin{align}
\mu &= \frac{3 (1+ w_d) (r_1 r_2 + 9)- r_1 r_2 (r_1+ r_2)}{r_1r_2 \Bigl(r_1+ r_2 - 3(1+ w_d)  \Bigr)- 27},\\
\alpha &= \frac{1}{3}\, \left[\frac{(r_1 r_2 - 9 w_d) (r_1+ r_2 - 3(1+ w_d))- 27}{r_1 r_2 \Bigl(r_1 + r_2 - 3 (1+ w_d)  \Bigr)- 27}\right].
\end{align}
Similar to the previous case, we also calculate the coupling parameters for the observational data OHD$+$SNe. In case of interacting quintessence (with $w_d = -0.98$), the coupling parameters are, $\mu = -0.074^{+0.129}_{-0.051}$ and $\alpha = 0.076^{+0.083}_{-0.119}$. For interacting vacuum DE, the values are,  $\mu= -0.060^{+0.138}_{-0.047}$ and $\alpha= 0.067^{+0.090}_{-0.126}$. For interacting phantom DE, the values are,  $\mu= -0.041^{+0.145}_{-0.056}$ and $\alpha= 0.055^{+0.089}_{-0.021}$. \\

3. Finally, we consider the possibility $\mu=\lambda$, that means, the interaction in  this case takes the form 
\begin{eqnarray}\label{int-C}
Q=3H\lambda \rho_t + 3H \alpha \rho^{\prime}_m,
\end{eqnarray}
for which the coupling parameters can be expressed as,

\begin{align}
\alpha &=\frac{1}{3} \left[\frac{r_1+r_2-3(1+w_d)-3}{r_1+r_2-3(1+w_d)}\right],\\
\mu & =\lambda =1-\frac{1}{3w_d} \left[\frac{r_1r_2}{r_1+r_2-3(1+w_d)}-3\right].
\end{align}
The coupling parameters for the combined analysis OHD$+$SNe
turn out to be 
$\alpha= -0.067^{+0.089}_{-0.157}$ and $\mu=\lambda= -0.106^{+0.149}_{-0.256}$ (interacting quintessence); 
$\alpha= -0.059^{+0.087}_{-0.171}$ and $\mu=\lambda= -0.088^{+0.149}_{-0.253}$ (interacting vacuum DE); 
$\alpha= -0.050^{+0.083}_{-0.145}$ and $\mu=\lambda= -0.063^{+0.143}_{-0.238}$ (interacting phantom DE).

From the above analysis it is observed that the mean values of the coupling parameters can be postive or negative depending on $w_d$, i.e., the nature of the dark energy . A change of sign in the interaction function means the change in the direction of the energy flow. So, an investigation of the evolution of $Q$ should be worthwhile. 

However, all phenomenological interactions do not have this property. For instance, the interactions of the form $Q = 3 H \lambda \rho_m$, or $Q = 3 H \mu \rho_d$ can be either positive or negative depending on the sign of the coupling parameters and cannot change the signature in the course of evolution. But, the interactions that are linear combinations of two energy densities $\rho_m$, $\rho_d$ and their derivaties, may have this feature. The interaction $Q$ can also be written down analytically. Inserting (\ref{rhom}), (\ref{rhod}) and the first derivative of (\ref{rhom}) into eqn. (\ref{interaction}) we find that, 

\begin{eqnarray}\label{interaction-explicit}
\bar{Q} = \frac{Q}{3H} = H_0^2\; F (z), 
\end{eqnarray}
where $F (z)$ determines the evolution of $\bar{Q}$ with $z$ and it is of the form 

\begin{eqnarray}\label{interaction-explicit-A}
F (z)= \Bigg[ A\; (1+z)^{r_1}+ B\; (1+z)^{r_2}\Bigg].
\end{eqnarray}

The constants $A$, $B$ are given by

\begin{align}
A &= \frac{\Omega_1}{w_d}\; \Bigg[\Bigl(\alpha \, r_1^2- \mu \left(3-r_1  \right)- \lambda\, r_1  \Bigr) + 3 (1+ w_d) \Bigl(\lambda- \alpha\, r_1  \Bigr)\Bigg], \\
B &= \frac{\Omega_2}{w_d}\; \Bigg[\Bigl(\alpha \, r_2^2- \mu \Bigl(3-r_2  \Bigr)- \lambda\, r_2\Bigr) + 3 (1+ w_d) \Bigl(\lambda- \alpha\, r_2  \Bigr)\Bigg],
\end{align}

Thus, it is clear that the interaction could change its sign if both of $A$ and $B$ are not of the same sign.  Further, one  may see that at some epoch there is a possibility for $\bar{Q}$ to attain its extremum, and it is found by the solution of the equation $d\bar{Q}/dz = 0$

\begin{eqnarray}
z_{ext} = -1 + \Bigg(-\frac{B r_2}{A r_1}\Bigg)^{\frac{1}{r_1-r_2}},
\end{eqnarray}
which could be real only with some conditions. Now, one can also try to see whether the extremum is a maximum or a minimum by calculating the second derivative 
\begin{align}
\frac{d^2 \bar{Q}}{dz^2}= H_0^2 \Bigl [A r_1 \left(r_1-1 \right) (1+z)^{r_1-2} \nonumber\\+ B r_2 \left(r_2-1 \right) (1+z)^{r_2-2} \Bigr],
\end{align}
at $z_{ext}$, the redshift at which $\bar{Q}$ may allow such possibilities. \\

In Figure \ref{figure-intABC}, we describe the behaviour of the interaction function (\ref{int-A}), (\ref{int-B}) and (\ref{int-C}) in the left, middle and right panels respectively while in each plot the solid, dot and dashed lines represent the interacting quintessence, interacting vacuum and the interacting phantom scenarios for the combined data OHD$+$SNe. From the plots we observe that the interactions given by equations (\ref{int-A}) and (\ref{int-C}) exhibit similar behaviour irrespective of the nature of dark energy while the model (\ref{int-B}) is completely different. For the models (\ref{int-A}) and  (\ref{int-C}), we find a smooth transition of $F (z)$ (i.e. $Q$) in the recent past (i.e. $z> 0$ but close to $0$) from its positive values ($Q >0$: energy flow takes place from DE to CDM) to negative values ($Q <0$: energy flow takes place from CDM to DE). That means the direction of energy flow changes during the evolution of the universe in presence of the interactions (\ref{int-A}) and (\ref{int-C}). On the other hand, for the model (\ref{int-B}), we find that $Q$ is always negative, that means, here, the energy flow always takes place from CDM to DE.

\begin{figure*}
\begin{center}
\includegraphics[angle=0, width=0.4\textwidth]{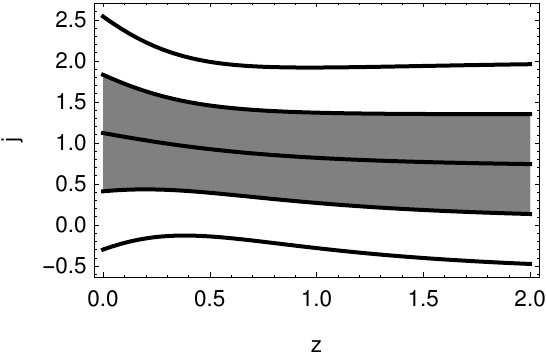}
\includegraphics[angle=0, width=0.38\textwidth]{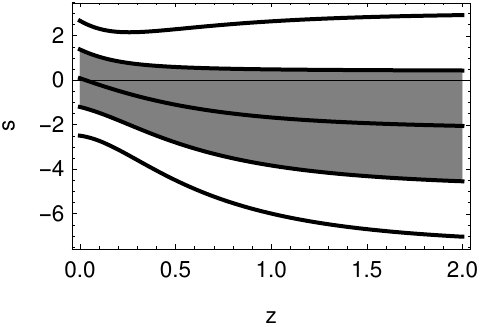}
\includegraphics[angle=0, width=0.38\textwidth]{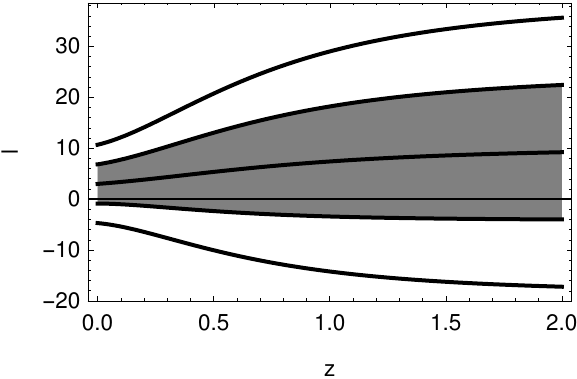}
\includegraphics[angle=0, width=0.38\textwidth]{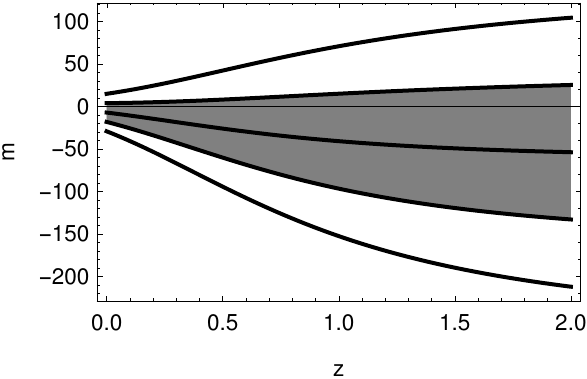}
\end{center}
\caption{\textit{The plots for the jerk $j$ (upper left), snap $s$ (upper right), lerk $l$ (bottom left) and $m$ parameter (bottom right) have been shown for the interacting quintessence scenario using the observational data OHD$+$SNe. In both panels we have shown the 1$\sigma$ ($68.3\%$) and 2$\sigma$ ($95.4\%$) confidence regions around the best fit curve (the central dark line).}}
\label{Fig-js-quintessence}
\end{figure*}
\begin{figure*}
\begin{center}
\includegraphics[angle=0, width=0.4\textwidth]{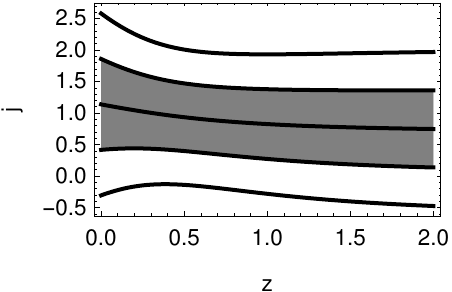}
\includegraphics[angle=0, width=0.38\textwidth]{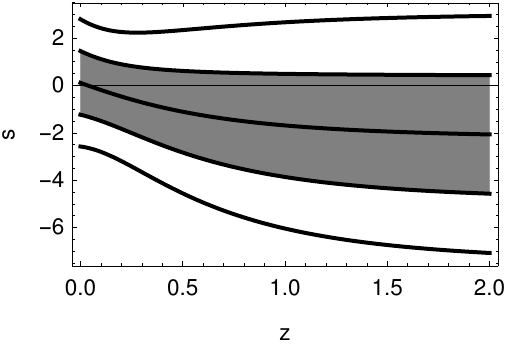}
\includegraphics[angle=0, width=0.37\textwidth]{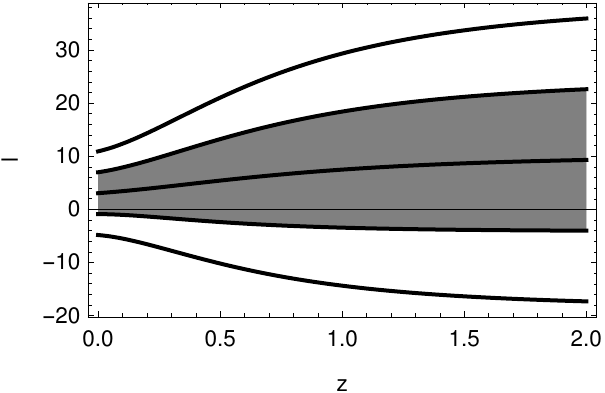}
\includegraphics[angle=0, width=0.37\textwidth]{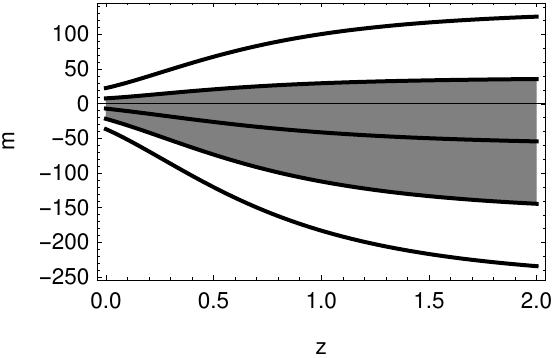}
\end{center}
\caption{\textit{The plots for the jerk $j$ (upper left), snap $s$ (upper right), lerk $l$ (bottom left) and $m$ parameter (bottom right) have been shown for the interacting vacuum scenario using the observational data OHD$+$SNe. In both panels we have shown the 1$\sigma$ ($68.3\%$) and 2$\sigma$ ($95.4\%$) confidence regions around the best fit curve (the central dark line).}}
\label{Fig-js-vacuum}
\end{figure*}
\begin{figure*}
\begin{center}
\includegraphics[angle=0, width=0.4\textwidth]{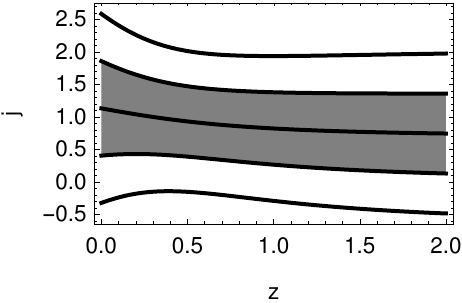}
\includegraphics[angle=0, width=0.39\textwidth]{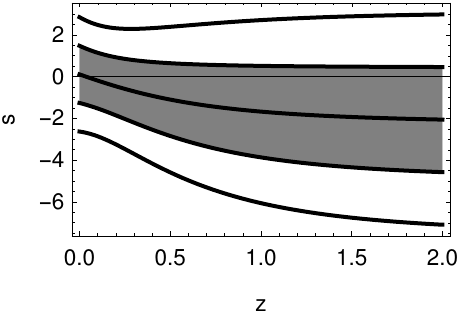}
\includegraphics[angle=0, width=0.39\textwidth]{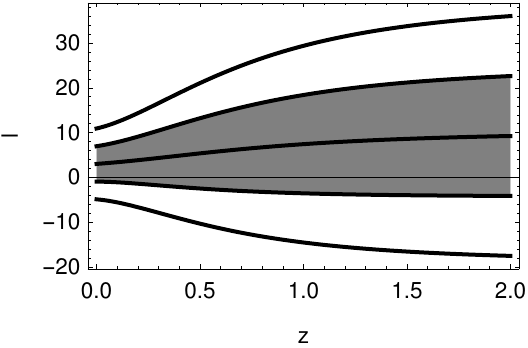}
\includegraphics[angle=0, width=0.39\textwidth]{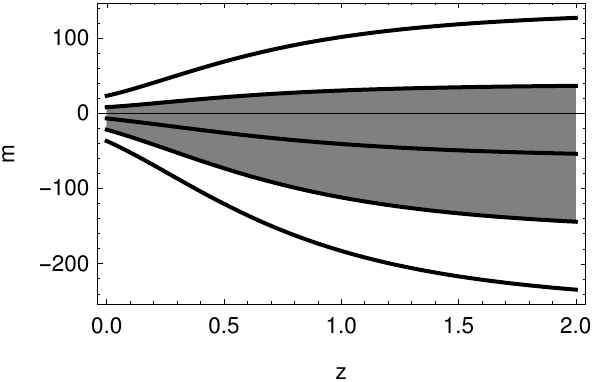}
\end{center}
\caption{\textit{The plots for the jerk $j$ (upper left), snap $s$ (upper right), lerk $l$ (bottom left) and $m$ parameter (bottom right) have been shown for the interacting phantom scenario using the observational data OHD$+$SNe. In both panels we have shown the 1$\sigma$ ($68.3\%$) and 2$\sigma$ ($95.4\%$) confidence regions around the best fit curve (the central dark line).}}
\label{Fig-js-phantom}
\end{figure*}
\begin{figure}
\begin{center}
\includegraphics[angle=0, width=0.4\textwidth]{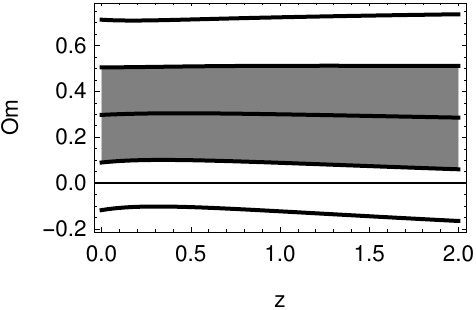}
\includegraphics[angle=0, width=0.4\textwidth]{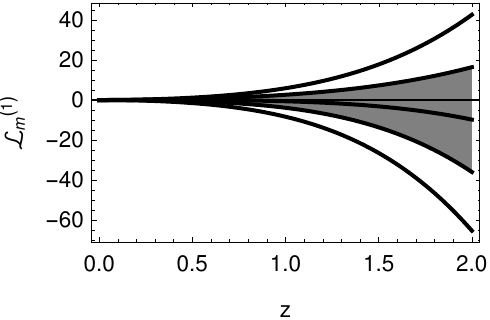}
\end{center}
\caption{\textit{In this figure for interacting quintessence model we show the variation of $Om$ diagnostic (upper panel) and its first derivative $\mathcal{L}_m^{(1)}$ (bottom panel)  for the observational data OHD$+$SNe. In both plots we further show the 1$\sigma$ ($68.3\%$) and 2$\sigma$ ($95.4\%$) confidence level plots around the best fit curve (the central dark line). }}
\label{Fig-om1}
\end{figure}
\begin{figure}
\begin{center}
\includegraphics[angle=0, width=0.4\textwidth]{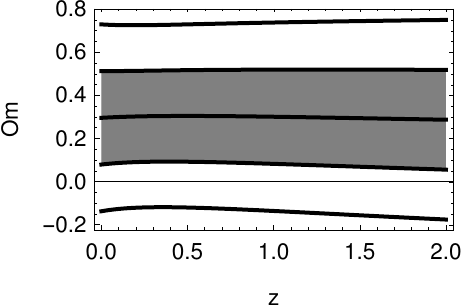}
\includegraphics[angle=0, width=0.4\textwidth]{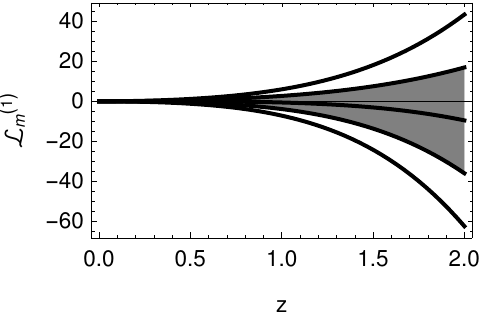}
\end{center}
\caption{\textit{For interacting vacuum model this figure shows the variation of $Om$ diagnostic (upper panel) and its first derivative $\mathcal{L}_m^{(1)}$ (bottom panel)  for the observational data OHD$+$SNe. In both plots we further show the 1$\sigma$ ($68.3\%$) and 2$\sigma$ ($95.4\%$) confidence level plots around the best fit curve (the central dark line).}}
\label{Fig-om2}
\end{figure}
\begin{figure}
\begin{center}
\includegraphics[angle=0, width=0.4\textwidth]{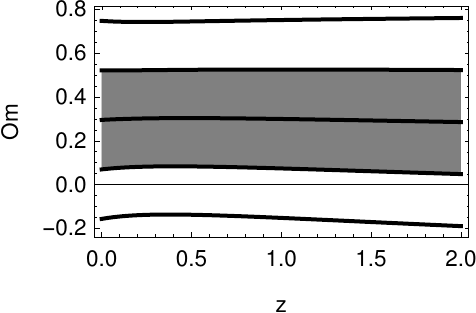}
\includegraphics[angle=0, width=0.4\textwidth]{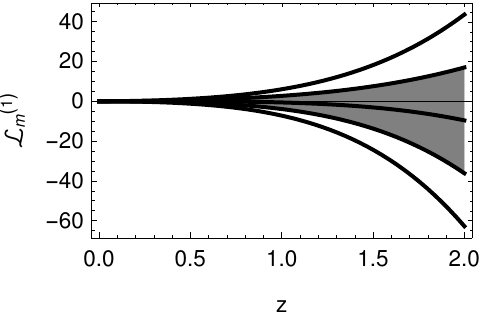}
\end{center}
\caption{\textit{In this figure for interacting phantom model we show the variation of $Om$ diagnostic (upper panel) and its first derivative $\mathcal{L}_m^{(1)}$ (bottom panel)  for the observational data OHD$+$SNe. As usual, in both plots we further show the 1$\sigma$ ($68.3\%$) and 2$\sigma$ ($95.4\%$) confidence level plots around the best fit curve (the central dark line).}}
\label{Fig-om3}
\end{figure}

\section{Interacting dark energy versus geometrical tests}
\label{sec-geometrical tests}

Cosmography \cite{Visser:2003vq} and $Om$ diagnostics \cite{SSS2008} are two well known geometrical tests that are generally used to differentiate several cosmological models by measuring their deviation from the $\Lambda$CDM cosmology. For the FLRW universe, its scale factor can be expanded as $a(t)= \sum _{i= 0}^{\infty} \frac{a^{(i)}(t_0)}{i!} (t-t_0)^i$, where $a^{(i)} (t_0)$ is the $i$-{th} derivative of $a(t)$ at $t= t_0$, and this Taylor series expansion provides some geometric quantities as coefficients, such as $H = \frac{1}{a}\frac{da}{dt}$, $q= -\frac{1}{aH^2}\frac{d^2a}{dt^2},$  
$j= \frac{1}{aH^3}\frac{d^3 a}{dt^3}$, $s= \frac{1}{aH^4}\frac{d^4a}{dt^4}$, $l= \frac{1}{aH^5}\frac{d^5a}{dt^5}$,  $m= \frac{1}{aH^6}\frac{d^6a}{dt^6}$, which are called the cosmographic parameters. The new quantities $j$, $s$, $l$, $m$ are known as the jerk, snap, lerk and $m$-parameter respectively. One may write down $j$, $s$, $l$, and $m$ in an alternative way

\begin{eqnarray}
j &=& (1+z) \frac{dq}{dz}+ q (1+2q),\label{j}\\
s &=& - (1+z)\frac{dj}{dz}- j (2+ 3q),\label{s}\\
l &=& - (1+z)\frac{ds}{dz}- s (3+ 4q),\label{l}\\
m &=& - (1+z)\frac{dl}{dz}- l (4+ 5q).\label{m}
\end{eqnarray}

Before the discovery of the accelerated expansion of the universe, $H$ indeed was an observational quantity and it is evolving. Its evolution is given by the next higher order derivative of the scale factor, namely $q$. Now that $q$ is an observable quantity and is found to evolve, so $j$ becomes the automatic choice of interest amongst these kinematical quantities \cite{ankan}. 

Now, in Figures \ref{Fig-js-quintessence},  \ref{Fig-js-vacuum}, \ref{Fig-js-phantom} we display the evolution of $j$ (upper left), $s$ (upper right), lerk (bottom left) and the $m$ parameter (bottom right) respectively, for interacting quintessence, interacting vacuum and interacting phantom scenarios using the observational data OHD$+$SNe with their $1\sigma$ and $2\sigma$ confidence levels around the best fit line (the central dark line of each plot).

We note that in all the three figures (i.e. Figs. \ref{Fig-js-quintessence},  \ref{Fig-js-vacuum} and \ref{Fig-js-phantom}), the present value of $j$, i.e. $j (z=0)$, is close to the corresponding value for the $\Lambda$CDM model, namely $j=1$. This indicates that around $z =0$, the interacting models resemble the $\Lambda$CDM model. The difference is more pronounced in the past history.

Let us introduce another geometric test known as $Om$ diagnostic, defined as \cite{SSS2008}

\begin{eqnarray}
Om (z)&=& \frac{h^2 (z)-1}{(1+z)^3-1},\label{om1}
\end{eqnarray}
where $h= H (z)/H_0$. This diagnostic only needs to know $H$, so it is a model independent diagnostic. It is evident that for a spatially flat $\Lambda$CDM, equation (\ref{om1}) takes the form $Om (z)= \Omega_{m0}$, irrespective of the redshift. That means, for any two distinct redshifts, say $z_i$ and $z_j$, $Om (z_i)- Om (z_j)= 0$ is the test for $\Lambda$CDM. Certainly, any deviation from this condition, a deviation from $\Lambda$CDM is indicated. 
Now, in the present model, where cold dark matter interacts with dark energy, we find that

\begin{align}
Om (z) & = \Omega_{1} \left[\frac{(1+z)^{r_1}-1}{(1+z)^3-1}\right]+ \left(1-\Omega_{1}\right ) \left[\frac{(1+z)^{r_2}-1}{(1+z)^3-1}\right].\label{om1b}
\end{align}
Thus, eqn. (\ref{om1b}) indicates that  $Om (z)$ is indeed different from $\Omega_{m0}$, and it  signals that the cosmological models do not mimic the $\Lambda$CDM model.

There is another diagnostic which is constructed from the first order derivative of the $Om$ diagnostic $\mathcal{L}^{(1)}_m$ as \cite{Seikel:2012cs}

\begin{equation}
\mathcal{L}_m^{1} = 3(1 +z)^2 (1-h^2)+ 2z (3+ 3z+ z^2)h h^{(1)},
\end{equation}
where $h^{(1)}= dh/dz$.  The model is like $\Lambda$CDM if $\mathcal{L}_m^{1} =0$ and $\mathcal{L}_m^{(1)} \neq 0$ implies that the model is not $\Lambda$CDM.

In Figures \ref{Fig-om1}, \ref{Fig-om2} and \ref{Fig-om3}, we show the behaviour of $Om$ (in the upper panel of all figures) and $\mathcal{L}_m^{1}$ (in the bottom panel of every figures) with the evolution of the universe in the context of interacting quintessence (Figure \ref{Fig-om1}), interacting vacuum (Figure \ref{Fig-om2}) and interacting phantom (Figure \ref{Fig-om3}) scenarios using the combined observational data OHD$+$SNe. We see that the evolution of $Om$ for all interacting dark energy models is almost same as that of the non-interacting $\Lambda$CDM and at present epoch ($z = 0$), the values of $Om$ for these models are close to $\Omega_{m0}$, the value of $Om$ taken by the $\Lambda$CDM model. On the other hand, although the evolutions of $\mathcal{L}_m^{1}$ (right panel of Figures \ref{Fig-om1}, \ref{Fig-om2}, \ref{Fig-om3}) are similar for all interacting models discussed here, but such evolutions significantly differ from the non-interacting $\Lambda$CDM at high redshifts. However, at low redshifts, the evolutions of $\mathcal{L}_m^{1}$ for all interacting models are same as that of the $\Lambda$CDM cosmology characterized by $\mathcal{L}_m^{1} =0$. 

\begin{figure*}
		\includegraphics[width=0.28\textwidth]{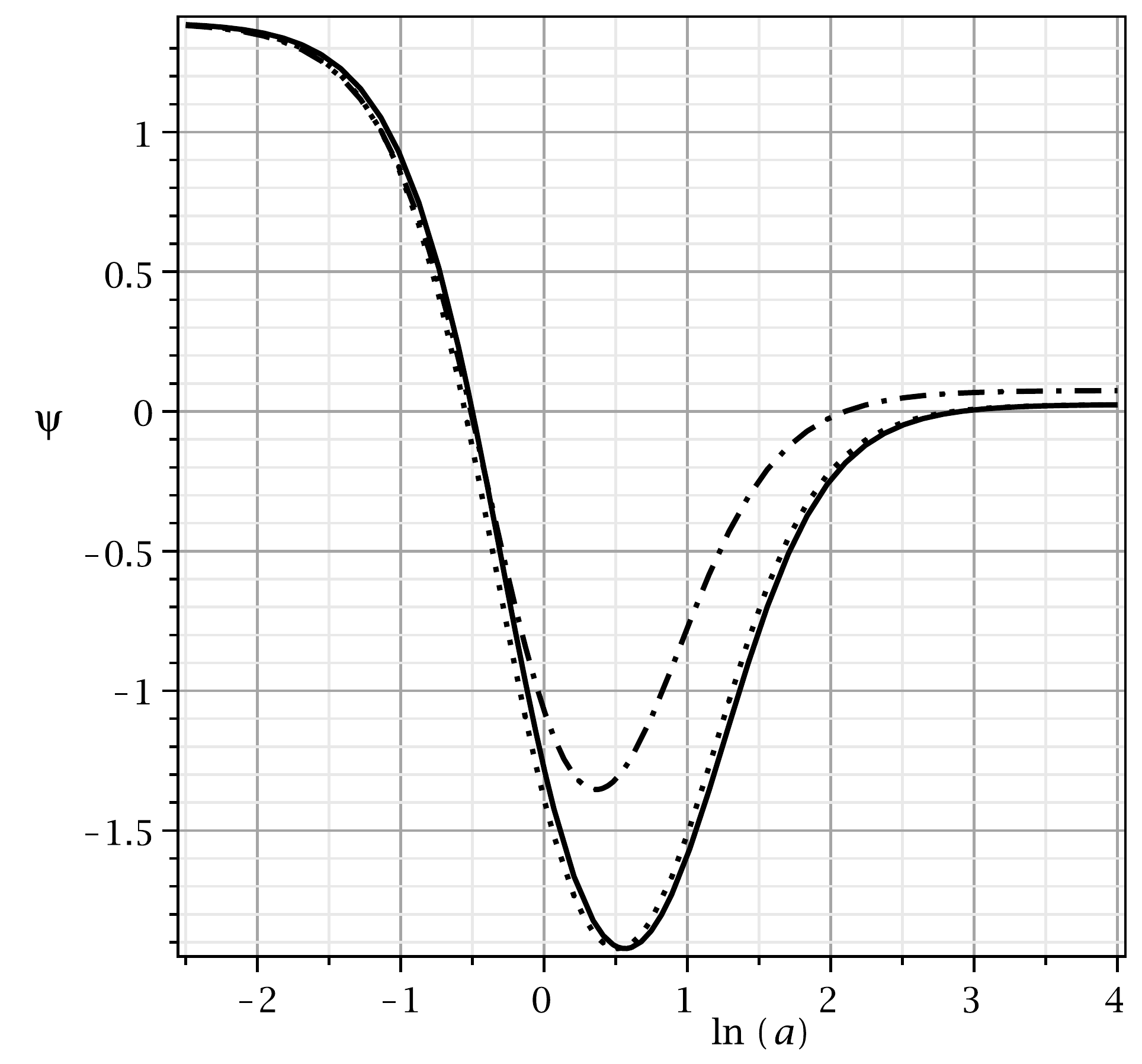}
		\includegraphics[width=0.28\textwidth]{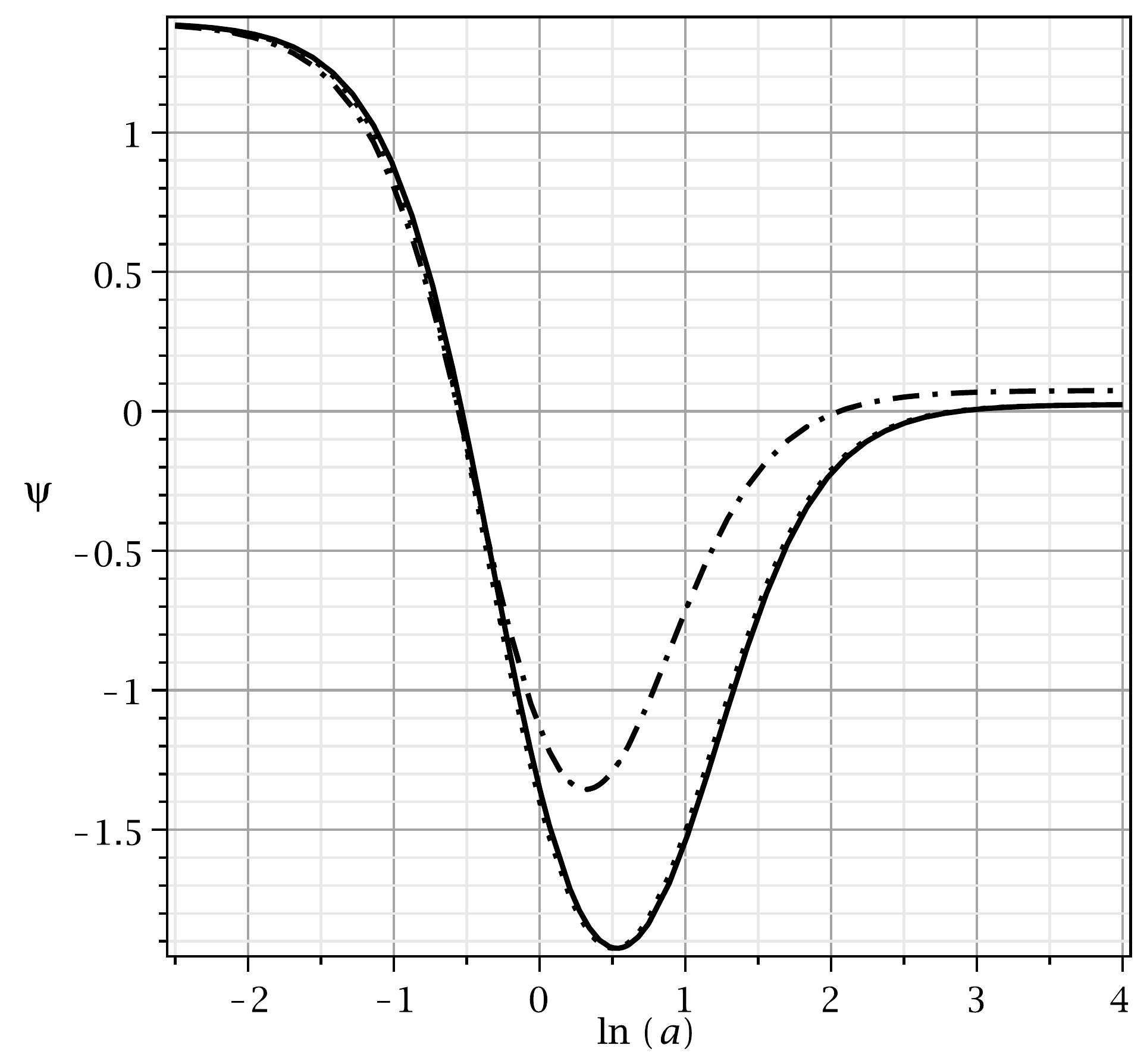}
		\includegraphics[width=0.28\textwidth]{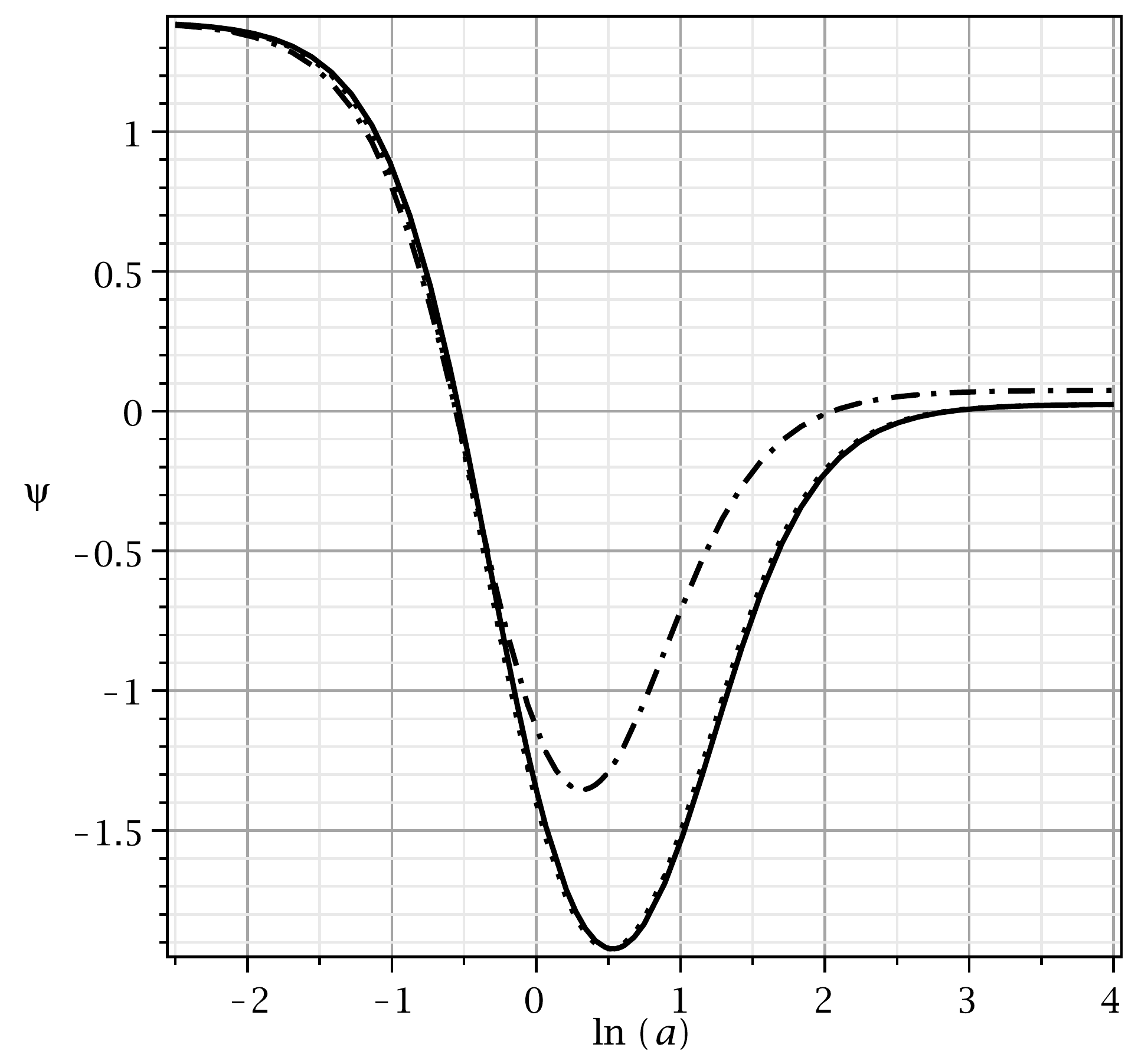}
	\caption{\textit{The figure shows the variation of the quantity $\psi$ determining the convex and concave nature of the entropy $S$ for the interacting cosmological models. The left, middle and right panel respectively stands for the interacting quintessence ($w_d =-0.98$), interacting vacuum ($w_d =-1$) and interacting phantom ($w_d= -1.02$) models. For the respective interacting scenario, we have used the parameters estimations ($r_1$, $\Omega_{d0}$) from the analysis OHD$+$SNe and we have taken the positive values of $r_2$ allowed in the same combined analysis. The solid, dot and dashdot curves in each plot respectively stand for $r_2 = 0.05, 0.1, 0.15$. }}
	\label{thermo_entropy_variation}
\end{figure*} 

\section{Thermodynamics of interacting dark energy}
\label{sec-thermo}

For an investigation of the thermodynamical properties of a cosmological model, one considers the universe as a thermodynamic system that is bounded by some cosmological horizon, and the matter content of the universe is enclosed within a volume defined by a radius not bigger than the cosmological horizon. It deserves mention that the idea of this horizon actually originates from the consideration of black hole thermodynamics. It has been shown that the thermodynamical properties which hold for a black hole are equally valid for a cosmological horizon \cite{gibbons, jacobson, paddy}. Furthermore, the first law of thermodynamics which holds in a black hole horizon can also be extracted from the first Friedmann equation in the FLRW universe when the universe is bounded by an apparent horizon. So, this gives a good motivation to choose the apparent horizon as the cosmological horizon to test the thermodynamic properties of any cosmological model and in the current section we have considered our universe to be bounded by the apparent horizon with radius $r_h= \left(H^2+ k/a^2  \right)^{-1/2}$ \cite{bak-rey} which, for $k= 0$, yields $r_h = 1/H$, the so-called Hubble horizon. 

However, according to the laws of thermodynamics, like any isolated macroscopic system, (i) the entropy should be non-decreasing with the expansion of the universe, that means if $S_f$, $S_h$ respectively stand for the entropy of the fluid and the entropy of the horizon containing the fluid, then the total entropy of the system, i.e. $S= S_f+ S_h$, should satisfy the relation $S^{\prime} \geq 0$, where prime denotes the differentiation with respect to $x= \ln a$,  and (ii) the entropy must be convex, that means $S^{\prime \prime} < 0$. The first criteria is the second law of thermodynamics and the final condition is required so as to ensure that if there is an extremum, it is a maximum. Now, considering the apparent horizon as the cosmolgical horizon, like in black hole physics, the entropy of the horizon $S_h$ can be taken as

\begin{eqnarray}\label{sp-thermo2}
S_h= \frac{k_B \mathcal{A}}{4\; l_{pl}^2} = 8 \pi^2 r_h^2.
\end{eqnarray} 
The final form results with choice of units as $\hbar= k_B= c= 8 \pi G = 1$; $k_B$, $\hbar$ are respectively the Boltzmann's constant and the Planck's constant respectively; $l_{pl} = (\sqrt{\hbar G/c^3})$ is the Planck's length; $\mathcal{A}= 4 \pi r_h^2$, is the area of the volume enclosed by the horizon radius. Furthermore, connecting with the black hole physics, the temperture of the apparent horizon is related with the horizon radius by the relation $T_h = 1/(2 \pi r_h)$, as given in a series of investigations \cite{jacobson, paddy, bak-rey, horizon-temperature-1, horizon-temperature-2,  horizon-temperature-3,  horizon-temperature-4, horizon-temperature-5}.

Now, if we denote $S_f = S_m + S_d$, where  $S_m$, $S_d$ being the entropies of the cold dark matter and the dark energy, and $T$ is the temperture of the composite matter (cold dark matter and dark energy) inside the horizon, then the first law of thermodynamics $ TdS = dE + p dV$ can be recast for the individual matter contents in the form 

\begin{align}
T dS_m &= dE_m + p_m dV  = dE_m,\label{sp-thermo4}\\
T d S_d &= dE_d + p_d dV,\label{sp-thermo5}
\end{align}
where $V= 4\pi r_h^3/ 3$, is the fluid volume; $E_m $, $E_d$ stand for the internal energies of the cold dark matter and dark energy given by $E_m = \frac{4}{3}\, \pi\, r_h^3\, \rho_m$ and $E_d = \frac{4}{3}\, \pi\, r_h^3\, \rho_d$ respectively. Now, differentiating equations (\ref{sp-thermo4}) and (\ref{sp-thermo5}) and (\ref{sp-thermo2}) with respect to the cosmic time one gets

\begin{align}
\left(\dot{S}_m,\, \dot{S}_d,\, \dot{S}_h \right) = \left(\frac{\dot{E}_m}{T},\,\,\,\frac{4 \, \pi \, p_d \, r_h^2 \,\dot{r}_h + \dot{E}_d}{T},\, 16 \pi^2 r_h \dot{r}_h   \right),
\end{align}

Using the above relations along with the asumption that the temperture of the fluid $T$ should be equal to that of the horizon temperture $T_h$, one can arrive at 

\begin{equation}\label{gslt}
\dot{S}= \dot{S}_m+ \dot{S}_d+ \dot{S}_h= 4 \pi^2 H r_h^6 \Bigl[\rho_m + (1+ w_d)\rho_d  \Bigr]^2
\end{equation}
which is of course positive. We should mention that the above relation has already been established in an interacting scenario where dark matter, dark energy and radiation are inteacting with each other \cite{manos}. In fact, the relation (\ref{gslt}) can be generalized for any number of interacting components. 

\par The equation (\ref{gslt}) takes the form

\begin{eqnarray}\label{gslt1.1}
S^\prime = \frac{16 \pi ^2}{H^4}\, \left( H^\prime \right)^2,
\end{eqnarray}
where the responsibility of the differentiation (denoted by a prime) has been transferred to $x=\ln  a$.
Differentiating this equation once more, one obtains

\begin{align}
S^{\prime \prime} = 2\,S^{\prime}\, \left( \frac{ H^{\prime \prime}}{H^{\prime}}- \frac{2\,H^{\prime}}{H} \right) = 2\,S^{\prime}\, \left( \frac{ h^{\prime \prime}}{h^{\prime}}- \frac{2\,h^{\prime}}{h} \right)= 2\,S^{\prime}\psi, 
\end{align}
where $h= H/H_0$ and $\psi= \left( \frac{ h^{\prime \prime}}{h^{\prime}}- \frac{2\,h^{\prime}}{h} \right)$. \\

For the interacting dark energy model we already found that $S^\prime \geq 0$, so the entropy function will be convex if $\psi < 0$. We now try to understand the behaviour of $S^{\prime \prime}$ with the evolution of the universe for the three interacting dark energy models discussed in this work. The behaviour of $S^{\prime \prime}$ (practically that of $\psi$ and $S^{\prime}$) is mainly dependent on the parameters $r_1$, $r_2$, $w_d$ and the density parameter for dark energy, $\Omega_{d0}$. We have already noted that, for all the models in the present work, for all observational data sets utilzed, the mean values of $r_1$ are positive and within $1\sigma$ confidence level, $r_1$ is strictly positive (see Table \ref{tableResult1}, Table \ref{tableResult2} and Table \ref{tableResult3}). On the other hand, the mean values of $r_2$ are negative for OHD and OHD$+$SNe but positive for SNe data, however, for all the observational data sets, $r_2$ is allowed to pick up both positive and negative values in the $2\sigma$ error bar. For the positive values of $r_2$ that are allowed in the combined analysis OHD$+$SNe, we have shown the variation of the quantity $\psi$ against $x = \ln a$, for all three interacting dark energy models, in Fig. \ref{thermo_entropy_variation}. From the figure we observe that close to the present epoch, the quantity $\psi$ is indeed negative, that means the entropy function is convex implying that the universe could be in thermodynamic equilibrium. On the other hand, for $r_2 < 0$, we end up with a singularity ($\psi \rightarrow -\infty$) at some finite scale factor `$a$'. That means the thermodynamical arguments seem to favour the region with $r_2 > 0$.

\section{Summary and Discussions}
\label{discu}

The current work presents an analytic description of an interacting dark sector and their cosmological consequences in a spatially flat FLRW universe. Both CDM and DE are assumed to have a constant equation of state parameter. DE and CDM interact amongst themselves and thus satisfy a common conservation equation. The rate of interaction has been taken to be a general function of the energy densities of the dark sectors (i.e. CDM and DE)  and their first order derivatives which recovers some well known and most used interactions in the literature as special cases. The evolution equations for CDM and DE can, in fact, be analytically solved. We remark that, if the linear interaction (\ref{int}) [or (\ref{interaction})] includes up to $n$-th order derivatives of the energy densities, then for some specific values of $n$, the evolution equations for CDM and DE can be analytically solved! This is the speciality of the linear interactions between both dark matter and dark energy when they assume constant equations of state.\\

The reconstructed interacting DE model has been studied for three different dark energy scenario, namely the quintessence, vacuum and phantom dark energy. For the interacting quintessence, the results summarized in Table \ref{tableResult1} indicate a small deviation from the $\Lambda$CDM cosmology. The total equation of state of this scenario shows a quintessential phase with a smooth transition from the decelerated expansion to current accelerated expansion at a recent past (upper panel of Figure \ref{quintqzwtot}).  Similarly interacting vacuum and the interacting phantom cases also indicate a slight deviation from the $\Lambda$CDM cosmology as observed in Table \ref{tableResult2} (interacting vacuum) and Table \ref{tableResult3} (interacting phantom). An interesting thing to note is that the ``total'' equation of state for all the cases are still in the quintessence regime within the $1\sigma$ confidence-level, although in a $2\sigma$ confidence-level, the phantom character of $w_{tot}$ is also not rejected. The transition from the decelerated phase to the current accelerted phase happens at $z < 1$, in all the cases, as suggested by observations. \\

It is shown that for some special cases of the interaction model, given by some choices of the coupling parameters appearing in the form of the interaction, the flow of energy might change its direction in the course of evolution. We found that if the interaction in the dark sector is characterized by any of the models in 
(\ref{int-A}) or  (\ref{int-C}), then a sign change in the interaction function happens from its positive values ($Q >0$, i.e. energy flows from DE to CDM) to negative ($Q <0$, i.e. energy flows from CDM to DE) and this is independent of $w_d$. However, 
for the model (\ref{int-B}), $Q$ is always negative, that means, here, the energy flow takes place from CDM to DE and this does not depend on the equation of state for DE, $w_d$.  \\

Further, we did an exercise of the geometric diagnostics for interacting dark energy. In all those cases, our analysis shows that at late time all the three interacting dark energy models are not too far from the $\Lambda$CDM cosmology, but, in no case the model reduces to the $\Lambda$CDM throughout the history of the evolution. \\

It deserves mention that the Akaike or the corrected Akaike Information Criteria does not pose any strong evidence against any of the three cases ($w_d >  -1$; $w_d = -1$ and $w_d < -1$) of dark energy considered with the interaction, but the Bayesian Information Criteria indeed poses evidence against all the three choices, although none of the models is actually ruled out completely. In this analysis, the base model is chosen to be the $\Lambda$CDM. \\

We also show that the interacting models (irrespective of the interaction function) satisfy the generalized second law of thermodynamics, i.e. the entropy should be non-decreasing. However, the analysis brings out some striking features depending on the sign of $r_2$. We found that for $r_2 >0$,  the universe can potentially reach the equilibrium only around the present epoch! For moderately high positive or negative values of $\ln a$,  the second derivative of the entropy, i.e. $S^{\prime\prime}$ has positive values. On the other hand, we found that for $r_2 <0$, the quantity $S^{\prime\prime}$ quantifying the equilibrium nature could be divergent to negative values at some finite scale factor. This might be considered to be a constraint on the interacting models which states that the positive values of $r_2$ give the most favorable scenarios. A negative value of $r_2$ implies the corresponding density, namely ${\rho}_{d}$, increases with evolution which is quite unphysical. So we see that both the consideration of density and the thermodynamic requirements are consistent with a positive $r_2$, which is allowed within a $1\sigma$ error bar for the present models.\\

One important point to note is the results indicate that when the different data sets are utilized, the $\Lambda$CDM is recovered in the $1\sigma$ confidence level. The departure from the $\Lambda$CDM is more pronounced in the past, i.e. at higher values of the redshift $z$ which is also supported from the geometrical tests. \\

The other intriguing thing is that there actually no clear indication in the results obtained that distinctly favours anyone amongst the three type of $w_{d}$ chosen, namely, $> -1$, $-1$ and $< -1.$

\section*{ACKNOWLEDGMENTS}
The authors thank the referee for some crucial clarifying comments. 
SP was supported by the National Post-Doctoral Fellowship (File No. PDF/2015/000640) from the Science and Engineering Research Board (SERB), Government of India. 



\begin{thebibliography}{}


\bibitem{Planck2015} Planck collaboration, Astron. Astrophys. \textbf{594}, A13 (2016), arXiv:1502.01589 [astro-ph.CO]


\bibitem{Weinberg1989} S. Weinberg,
Rev. Mod. Phys. {\bf 61}, 1 (1989)

\bibitem{paddy} T. Padmanabhan, Phys. Rept. {\bf 380}, 235 (2003), arXiv:hep-th/0212290

\bibitem{CDS1998}R. R. Caldwell, R. Dave and P. J. Steinhardt, Phys. Rev. Lett. \textbf{80} 1582 (1998),  arXiv:astro-ph/9708069 

 
\bibitem{varun} V. Sahni and A. A. Starobinsky, Int. J. Mod. Phys. D {\bf 9}, 373 (2000), arXiv:astro-ph/9904398
 
\bibitem{CST} E. J. Copeland, M. Sami and S. Tsujikawa, 
Int. J. Mod. Phys. D. \textbf{15}, 1753 (2006), arXiv:hep-th/0603057


\bibitem{AT} L. Amendola and S. Tsujikawa, Dark Energy: Theory and Observations,  Cambridge University Press, Cambridge, UK (2010)



\bibitem{Bamba:2012cp} 
  K.~Bamba, S.~Capozziello, S.~Nojiri and S.~D.~Odintsov,
  Astrophys.\ Space Sci.\  {\bf 342}, 155 (2012),
  arXiv:1205.3421 [gr-qc].



\bibitem{Zlatev} I. Zlatev, L. Wang and P. J. Steinhardt, Phys. Rev. Lett. \textbf{82}, 896 (1999), arXiv:astro-ph/9807002

\bibitem{Wetterich-ide1} C. Wetterich,
Astron. Astrophys. \textbf{301}, 321 (1995), arXiv:hep-th/9408025

\bibitem{Amendola-ide1} L. Amendola,
Phys. Rev. D \textbf{62}, 043511 (2000), arXiv:astro-ph/9908023

\bibitem{Amendola-ide2} L. Amendola and C. Quercellini, Phys. Rev. D \textbf{68}, 023514 (2003), arXiv:astro-ph/0303228



\bibitem{Pavon:2005yx} 
D.~Pav\'{o}n and W.~Zimdahl,
Phys.\ Lett.\ B {\bf 628}, 206 (2005),
arXiv:gr-qc/0505020




\bibitem{delCampo:2008sr} 
S.~del Campo, R.~Herrera and D.~Pav\'{o}n,
Phys.\ Rev.\ D {\bf 78}, 021302 (2008),
arXiv:0806.2116 [astro-ph]




\bibitem{delCampo:2008jx} 
S.~del Campo, R.~Herrera and D.~Pav\'{o}n,
JCAP {\bf 0901}, 020 (2009),
arXiv:0812.2210 [gr-qc]


\bibitem{Billyard-ide1} A. P. Billyard and A. A. Coley, Phys. Rev. D \textbf{61}, 083503 (2000), arXiv:astro-ph/9908224

\bibitem{Zimdahl-ide1} W. Zimdahl, D. Pav\'{o}n and L. P. Chimento, Phys. Lett. B \textbf{521}, 133 (2001), arXiv:astro-ph/0105479



\bibitem{Herrera-ide1} R. Herrera, D. Pav\'{o}n and W. Zimdahl,
Gen. Relt. Grav. \textbf{36}, 2161 (2004), arXiv:astro-ph/0404086

\bibitem{Chimento-ide1} L. P. Chimento, A. S. Jakubi, D. Pav\'{o}n and W. Zimdahl,
Phys. Rev. D \textbf{67}, 083513 (2003), arXiv:astro-ph/0303145



\bibitem{He1} J.-H. He and B. Wang,
JCAP \textbf{0806}, 010 (2008), arXiv:0801.4233 [astro-ph]


\bibitem{Quartin1} M.~Quartin, M.~O.~Calvao, S.~E.~Joras, R.~R.~R.~Reis and I.~Waga,
JCAP \textbf{0805}, 007 (2008), arXiv:0802.0546 [astro-ph]


\bibitem{Maartens2008} C. G. Boehmer, G. Caldera-Cabral, R. Lazkoz and R. Maartens,
Phys. Rev. D \textbf{78}, 023505 (2008), arXiv:0801.1565 [gr-qc]


\bibitem{Maartens2009} G. Caldera-Cabral, R. Maartens and L. A. Urena-Lopez,
Phys. Rev. D \textbf{79}, 063518 (2009), arXiv:0812.1827 [gr-qc]



\bibitem{Valiviita:2009nu}
J.~Valiviita, R.~Maartens and E.~Majerotto,
Mon.\ Not.\ Roy.\ Astron.\ Soc.\  {\bf 402}, 2355 (2010),
arXiv:0907.4987 [astro-ph.CO]




\bibitem{Chimento2010} L. P. Chimento,
Phys. Rev. D \textbf{81}, 043525 (2010), arXiv:0911.5687 [astro-ph.CO]



\bibitem{Clemson:2011an}
T.~Clemson, K.~Koyama, G.~B.~Zhao, R.~Maartens and J.~Valiviita,
Phys.\ Rev.\ D {\bf 85}, 043007 (2012),
arXiv:1109.6234 [astro-ph.CO]




\bibitem{Chen:2011cy} 
  X.~m.~Chen, Y.~Gong, E.~N.~Saridakis and Y.~Gong,
  Int.\ J.\ Theor.\ Phys.\  {\bf 53}, 469 (2014),
  arXiv:1111.6743 [astro-ph.CO]
  

\bibitem{Pan2013} S. Pan and S. Chakraborty, Eur. Phys. J. C \textbf{73}, 2575 (2013), arXiv:1303.5602 [gr-qc]



\bibitem{Faraoni:2014vra} 
  V.~Faraoni, J.~B.~Dent and E.~N.~Saridakis,
  Phys.\ Rev.\ D {\bf 90}, 063510 (2014),
  arXiv:1405.7288 [gr-qc]
  



\bibitem{Pan:2012ki} 
S.~Pan, S.~Bhattacharya and S.~Chakraborty,
Mon.\ Not.\ Roy.\ Astron.\ Soc.\  {\bf 452}, 3038 (2015), arXiv:1210.0396 [gr-qc]

\bibitem{AP} A. Paliathanasis and M. Tsamparlis, Phys. Rev. D
\textbf{90}, 043529 (2014),  arXiv:1408.1798 [gr-qc]




\bibitem{Duniya:2015nva}
  D.~G.~A.~Duniya, D.~Bertacca and R.~Maartens,
  Phys.\ Rev.\ D {\bf 91}, 063530 (2015),
  arXiv:1502.06424 [astro-ph.CO]
  



\bibitem{Valiviita:2015dfa}
  J.~Valiviita and E.~Palmgren,
  JCAP {\bf 1507}, 015 (2015),
  arXiv:1504.02464 [astro-ph.CO]
  
  
\bibitem{delCampo:2015vha} 
  S.~del Campo, R.~Herrera and D.~Pav\'{o}n,
  Phys.\ Rev.\ D {\bf 91},  123539 (2015),
  arXiv:1507.00187 [gr-qc].  

\bibitem{Pan:2016ngu} 
S.~Pan and G.~S.~Sharov, Mon. Not. Roy. Astron. Soc. \textbf{472}, 4736 (2017),
arXiv:1609.02287 [gr-qc]




\bibitem{Mukherjee:2016shl} 
A.~Mukherjee and N.~Banerjee, Class. Quant. Grav. \textbf{34}, 035016  (2017),
arXiv:1610.04419 [astro-ph.CO]


\bibitem{Pan2017} G. S. Sharov, S. Bhattacharya, S. Pan, R. C. Nunes and S. Chakraborty, Mon.\ Not.\ Roy.\ Astron.\ Soc.\ \textbf{466}, 3497 (2017), arXiv:1701.00780 [gr-qc]


\bibitem{Odintsov:2017icc} 
  S.~D.~Odintsov, V.~K.~Oikonomou and P.~V.~Tretyakov,
  Phys.\ Rev.\ D {\bf 96},  044022 (2017),
  arXiv:1707.08661 [gr-qc].

\bibitem{Yang:2017yme} 
  W.~Yang, N.~Banerjee and S.~Pan,
  Phys.\ Rev.\ D {\bf 95}, 123527 (2017),
  arXiv:1705.09278 [astro-ph.CO]
  

\bibitem{DiValentino:2017iww} 
  E.~Di Valentino, A.~Melchiorri and O.~Mena, Phys. Rev. D \textbf{96}, 043503 (2017),
  arXiv:1704.08342 [astro-ph.CO]  

\bibitem{Yang:2017zjs} 
  W.~Yang, S.~Pan and J.~D.~Barrow,
  Phys.\ Rev.\ D {\bf 97},  043529 (2018),
  [arXiv:1706.04953 [astro-ph.CO]]
  
\bibitem{Yang:2017ccc} 
   W.~Yang, S.~Pan and D.~F.~Mota,
  Phys.\ Rev.\ D {\bf 96},  123508 (2017),
  [arXiv:1709.00006 [astro-ph.CO]]
  

\bibitem{Santos:2017bqm} 
  L.~Santos, W.~Zhao, E.~G.~M.~Ferreira and J.~Quintin,
  arXiv:1707.06827 [astro-ph.CO]

\bibitem{Kumar:2017bpv} 
  S.~Kumar and R.~C.~Nunes,
  arXiv:1709.02384 [astro-ph.CO]
  


\bibitem{Salvatelli2014} V. Salvatelli, N. Said,  M. Bruni,  A. Melchiorri,  D. Wands, Phys. Rev. Lett. \textbf{113}, 181301 (2014), arXiv:1406.7297 [astro-ph.CO]

\bibitem{Weiqiang1} W. Yang W and L. Xu, Phys. Rev. D \textbf{89}, 083517 (2014), arXiv:1401.1286 [astro-ph.CO]

\bibitem{Weiqiang2}  W. Yang W and L. Xu, Phys. Rev. D \textbf{90}, 083532 (2014), arXiv:1409.5533 [astro-ph.CO]


\bibitem{yang:2014vza}
W.~Yang and L.~Xu,
JCAP {\bf 1408}, 034 (2014),
arXiv:1401.5177 [astro-ph.CO]


\bibitem{Nunes:2016dlj} 
R.~C.~Nunes, S.~Pan and E.~N.~Saridakis,
Phys.\ Rev.\ D {\bf 94}, 023508 (2016), arXiv:1605.01712 [astro-ph.CO]




\bibitem{Kumar:2016zpg} 
S.~Kumar and R.~C.~Nunes,
Phys. Rev. D \textbf{94}, 123511 (2016),
arXiv:1608.02454 [astro-ph.CO]


\bibitem{Bruck2016}  C. Bruck, J. Mifsud and J. Morrice, arXiv:1609.09855 [astro-ph.CO]



\bibitem{Weiqiang3} W. Yang, H. Li, Y. Wu and J. Lu, JCAP \textbf{10}, 007 (2016), arXiv:1608.07039 [astro-ph.CO]


\bibitem{PTram2016} A. Pourtsidou and T. Tram, Phys. Rev. D \textbf{94}, 043518 (2016), arXiv:1604.04222 [astro-ph.CO]




\bibitem{Sadjadi:2006qb} 
H.~M.~Sadjadi and M.~Honardoost,
Phys.\ Lett.\ B {\bf 647}, 231 (2007),
arXiv:gr-qc/0609076




\bibitem{Pan:2014afa} 
S.~Pan and S.~Chakraborty,
Int.\ J.\ Mod.\ Phys.\ D {\bf 23}, 1450092 (2014),
arXiv:1410.8281 [gr-qc]



\bibitem{Carroll} S. M. Carroll, M. Hoffman and M. Trodden, 
Phys. Rev. D \textbf{68}, 023509 (2003),  astro-ph/0301273

\bibitem{Cline} J. M. Cline, S. Jeon and G. D. Moore, Phys. Rev. D \textbf{70}, 043543 (2004), arXiv:hep-ph/0311312

\bibitem{Wang:2016lxa} 
  B.~Wang, E.~Abdalla, F.~Atrio-Barandela and D.~Pavon,
  Rept.\ Prog.\ Phys.\  {\bf 79}, 096901 (2016),
  arXiv:1603.08299 [astro-ph.CO].
  
 \bibitem{ForemanMackey:2012ig} 
  D.~Foreman-Mackey, D.~W.~Hogg, D.~Lang and J.~Goodman,
  Publ.\ Astron.\ Soc.\ Pac.\  {\bf 125}, 306 (2013),
  [arXiv:1202.3665 [astro-ph.IM]]
  
  
  \bibitem{betoule} M.~Betoule {\it et al.} [SDSS Collaboration], Astron. Astrophys. {\bf 568}, A22 (2014), arXiv:1401.4064 [astro-ph.CO]

\bibitem{farooqmaniaratra} O. Farooq, D. Mania and B. Ratra, Astrophys. J. {\bf 764}, 138 (2013), arXiv:1211.4253 [astro-ph.CO]

\bibitem{simon} J.~Simon, L.~Verde and R.~Jimenez, Phys. Rev. D {\bf 71}, 123001 (2005), arXiv:astro-ph/0412269

\bibitem{stern} D.~Stern, R.~Jimenez, L.~Verde, M.~Kamionkowski and S.~A.~Stanford, JCAP {\bf 02}, 008 (2010), arXiv:0907.3149 [astro-ph.CO]

\bibitem{zhang} C. Zhang, H. Zhang, S. Yuan, S.-Q. Liu, T.-J. Zhang, Y.-C. Sun, Research in Astronomy and Astrophysics {\bf 14}, 1221 (2014), arXiv:1207.4541 [astro-ph.CO]

\bibitem{moresco} M.~Moresco, L.~Verde, L.~Pozzetti, R.~Jimenez and A.~Cimatti, JCAP {\bf 07}, 053 (2012), arXiv:1201.6658 [astro-ph.CO]

\bibitem{delubac} T.~Delubac {\it et al.} [BOSS Collaboration], Astron. Astrophys. {\bf 574}, A59 (2015), arXiv:1404.1801 [astro-ph.CO]



\bibitem{aic} H. Akaike,
IEEE Transactions on
Automatic Control. \textbf{19}, 716 (1974)

\bibitem{bic} G. Schwarz,
Ann. Statist. \textbf{6}, 461 (1978)



\bibitem{aicc} N. Sugiura,
Communications in Statistics - Theory and Methods, \textbf{A7}, 13 (1978)



\bibitem{Liddle} A. R. Liddle,  
Mon. Not. Roy. Astron. Soc. \textbf{377}, L74 (2007), arXiv:astro-ph/0701113 



\bibitem{Visser:2003vq} 
  M.~Visser, Class.\ Quant.\ Grav.\  {\bf 21}, 2603 (2004), arXiv:gr-qc/0309109
  
 
\bibitem{SSS2008} V. Sahni, A. Shafieloo, and A. A. Starobinsky,  
Phys. Rev. D. {\bf 78}, 103502, 2008,  arXiv:0807.3548 [astro-ph]  
  
\bibitem{ankan} A. Mukherjee and N. Banerjee, Phys. Rev. D \textbf{93}, 043002 (2016), arXiv:1601.05172 [gr-qc]



\bibitem{Seikel:2012cs} 
M.~Seikel, S.~Yahya, R.~Maartens and C.~Clarkson,
Phys.\ Rev.\ D {\bf 86}, 083001 (2012),
arXiv:1205.3431 [astro-ph.CO]

\bibitem{gibbons} G. W. Gibbons and S. W. Hawking, Phys. Rev. D
\textbf{15}, 2738 (1977)

\bibitem{jacobson} T. Jacobson, Phys. Rev. Lett. \textbf{75}, 1260 (1995), arXiv:gr-qc/9504004



\bibitem{bak-rey}  D. Bak and S. J. Rey, Class. Quant. Grav. \textbf{17}, L83 (2000), arXiv:hep-th/9902173



\bibitem{horizon-temperature-1} A. V. Frolov and L. Kofman, JCAP \textbf{0305}, 009 (2003), arXiv:hep-th/0212327

\bibitem{horizon-temperature-2} U. H. Danielsson, Phys. Rev. D \textbf{71}, 023516 (2005), arXiv:hep-th/0411172

\bibitem{horizon-temperature-3} R. Bousso, Phys. Rev. D \textbf{71}, 064024 (2005), arXiv:hep-th/0412197

\bibitem{horizon-temperature-4} R. G. Cai and S. P. Kim, JHEP \textbf{0502}, 050 (2005), arXiv:hep-th/0501055
 
\bibitem{horizon-temperature-5} M. Akbar and R. G. Cai, Phys. Rev. D \textbf{75}, 084003 (2007), arXiv:hep-th/0609128 

\bibitem{manos}M.~Jamil, E.~N.~Saridakis and M.~R.~Setare, JCAP {\bf 1011}, 032 (2010), arXiv:1003.0876 [hep-th]
 

\end{thebibliography}
\end{document}